\documentclass[prb,twocolumn,showpacs,amsmath,amssymb]{revtex4-1}

\usepackage[pdftex]{graphicx}
\usepackage{ifthen}
\usepackage{color} 
\usepackage{bm} 
\usepackage[normalem]{ulem}
\usepackage{multirow}
\usepackage{hyperref}
\usepackage{bbold}
\hypersetup{
 pdfnewwindow=true, colorlinks=true,
 linkcolor=blue, anchorcolor=blue,
 citecolor=blue, filecolor=blue,
 menucolor=blue, urlcolor=blue}
\usepackage{breakurl}

\def\ang{\theta}
\def\t{\bm{t}}
\def\q{\bm{q}}
\def\r{\bm{r}}
\def\k{\bm{k}}
\def\G{\bm{G}}
\def\kp{\bm{k'}}
\def\Gp{\bm{G'}}
\def\kpp{\bm{k''}}
\def\Gpp{\bm{G''}}
\def\R{\bm{R}}
\def\P{\bm{P}}
\def\kap{\bm{\kappa}}
\def\dsum{\displaystyle\sum}
\def\ket#1{\vert #1 \rangle}
\def\bra#1{\langle #1 \vert}
\def\me#1#2#3{\bra{#1}#2\ket{#3}}
\def\dpr#1#2{\langle #1 \vert #2 \rangle} 

\begin{document}

\title{Theory of the Raman spectrum of rotated double-layer graphene}

\author{Sinisa Coh}
\email{sinisa@civet.berkeley.edu} 
\author{Liang Z. Tan}
\author{Steven G. Louie} 
\author{Marvin L. Cohen} 
\affiliation{Department of Physics, University of California at
  Berkeley, Berkeley, CA 94720, USA}
\affiliation{Materials Sciences Division, Lawrence Berkeley National
  Laboratory, Berkeley, CA 94720, USA}

\date{\today}

\pacs{  78.67.Wj, 73.22.Pr, 63.22.Rc, 78.30.Na }

\begin{abstract}
  We study theoretically the Raman spectrum of the rotated
  double-layer graphene, consisting of two graphene layers rotated
  with respect to each other by an arbitrary angle $\theta$. We find a
  relatively simple dependence of the Raman G peak intensity on the
  angle $\theta$. On the other hand, the Raman 2D peak position,
  intensity, and width show a much more complicated dependence on the
  angle $\theta$. We account for all of these effects, including
  dependence on the incoming photon energy, in good agreement with the
  experimental data. We find that it is sufficient to include the
  interaction between the graphene layers on the electronic degrees of
  freedom (resulting in the occurrence of Van Hove singularities in
  the density of states). We assume that the phonon degrees of freedom
  are unaffected by the interaction between the layers.  Furthermore,
  we decompose the Raman 2D peak into two components having very
  different linewidths; these widths are almost independent of the
  angle $\theta$. The change in the intensity and the peak position of
  one of these two components gives insight into the dependence of the
  overall Raman 2D peak features as a function of the angle $\theta$.
  Furthermore, we study the influence of the coherence on the Raman
  signal, and we separately study the influence of the interaction
  between the layers on the electron wavefunctions and energies.
  Additionally, we show regions in the phonon spectrum giving rise to
  the Raman 2D peak signal.  This work provides an insight into the
  interplay between the mechanical degree of freedom (angle $\theta$)
  and the electronic degrees of freedom (singularities in the density
  of states) in rotated double-layer graphene.  Additionally, this
  work provides a way to establish experimentally the value of the
  rotation angle $\theta$ using Raman spectroscopy measurement. This
  procedure becomes even more robust if one repeats the Raman
  spectroscopy measurement with a different incoming photon energy.
\end{abstract}

\maketitle

\section{Introduction}
\label{sec:intro}

The electronic band structure of a single graphene layer near the
Fermi level consists of Dirac-cone like structure at the Brillouin
zone edge (K point). In this work we study the rotated double-layer
graphene (also refereed to as the twisted bilayer graphene) which
consists of two single-layers of graphene that are rotated with
respect to each other by an arbitrary angle $\ang$. In the special
case when $\ang=0^{\circ}$ the Dirac cones from the two layers are
exactly on top of each other in reciprocal space.  However, rotation
of one of the graphene layers in real space ($\ang\neq0^{\circ}$) is
accompanied by a corresponding rotation of its band structure in
reciprocal space (around the origin of the reciprocal space).
Therefore, when $\ang\neq0^{\circ}$ Dirac cones of the two graphene
layers are no longer on top of each other in reciprocal space, but are
separated, proportionally to $\sim\sin\theta/2$. Nevertheless the two
Dirac cones are still overlapping in a small region of reciprocal
space in between the cones. From a perturbation theory argument, one
would expect that the interaction between the Dirac cones of the two
graphene layers will be particularly strong in region where the Dirac
cones are overlapping.  Indeed, interaction between the layers in the
overlap region opens a hybridization gap and leads to Van Hove
singularities in the density of states of the rotated double-layer
graphene. Since the position of the overlap region depends on the
angle $\ang$, we expect that the rotated double-layer graphene will
have an interesting coupling between the mechanical degree of freedom
(angle $\ang$) and the electronic degrees of freedom (singularities in
the density of states). Many interesting properties of rotated
double-layer graphene arise from this tunability, and they have
recently been attracting a lot of interest\cite{pap01, pap02, pap03,
  pap04, Bistritzer2011, pap06, pap07, pap08, pap09, pap10, pap11, pap12,
  kim2012, pap38}.

In this work we study theoretically the influence of the angle $\ang$
on the Raman spectrum of rotated double-layer graphene. Raman
spectroscopy is an experimental technique commonly used to
characterize carbon based materials as discussed in detail in
Ref.~\onlinecite{pap17}.  Since Raman spectroscopy uses incoming
photons with a well defined energy, one can use this spectroscopy to
study selectively certain regions of the electronic density of states.
Therefore, we can expect an interesting dependence of the Raman signal
of the rotated double-layer graphene as the angle $\ang$ is varied.
Such a dependence of the Raman signal was demonstrated in some recent
experimental
studies\cite{pap21,pap22,pap23,pap24,pap25,kim2012,pap31,pap38}.

The two most prominent Raman signals in graphene based systems are
Raman G peak (close to 1600~cm$^{-1}$) and Raman 2D (or G$'$) peak
(close to 2700~cm$^{-1}$). The Raman G peak in the graphene based
systems is a simpler process than the 2D peak since it involves
creation of just one phonon per each scattered photon. Considering
momentum conservation and assuming a negligible momentum of the photon
we conclude that the created phonon must occur at the Brillouin zone
center. On the other hand, Raman 2D peak involves emission of two
phonons per each scattered photon. In this case the momentum
conservation implies that the two emitted phonons have arbitrary but
opposite momenta.  In this work we study both of these Raman peaks (G
and 2D) in the rotated double-layer graphene as a function of angle
$\ang$. We find a relatively simple dependence of the Raman G peak on
the angle $\ang$. Namely, when the incoming photon energy is
comparable to the separation between the Van Hove singularities of the
rotated double-layer graphene, there is a significant increase
($\sim$70) in the G peak intensity. On the other hand, the Raman 2D
peak intensity, position, and width show a much more complicated
dependence on the angle $\ang$. All of these features, including the
incoming light frequency dependence, are well reproduced in our
calculation and agree well with experimental data (detailed comparison
is shown in Ref.~\onlinecite{kim2012}).  Furthermore, these results
provide a simple way to experimentally determine the angle $\ang$ of a
rotated double-layer graphene. The angle determination procedure
becomes even more robust if one performs Raman spectroscopy with two
(or more) different incoming photon energies.

We compute the Raman spectra of the rotated double-layer graphene
using a super-cell tight-binding method. Additionally, for the Raman
2D peak we confirm our findings using a continuum model method. In the
super-cell method we choose special values of $\ang$ for which there
exists a super-periodicity between the two graphene layers. In the
super-cell method we treat this enlarged commensurate super-cell as a
unit cell of our system.  On the other hand, in the continuum model
calculation we rely on a simple Dirac equation description of a
single-layer graphene and we add interaction with the other layer in
the restricted Hilbert space.  These continuum model calculations are
less numerically demanding than the super-cell calculations, since
they do not rely on the super-periodicity between the two graphene
layers. However, we expect that the super-cell tight-binding method is
more reliable, and we find that it compares better with experimental
data. Unless explicitly mentioned, the results reported here refer to
the super-cell tight-binding method.

We provide details of both the super-cell tight-binding method and the
continuum model method in Sec.~\ref{sec:methods}. In
Sec.~\ref{sec:results} we present results of the Raman G and 2D peaks
calculations in the rotated double-layer graphene case as a function
of the angle $\ang$. We also provide a detailed analysis of these
Raman peaks in the rotated double-layer graphene.

\section{Methods}
\label{sec:methods}

In the Raman process the incoming photon creates a virtual
electron-hole pair which then emits (or in some cases, absorbs) a
phonon excitation quantum (or a quantum of some other excitation). In
the first-order Raman process, the interaction of the single incoming
photon results in the emission of a single phonon excitation. In the
second-order Raman process two phonons are emitted for each
interaction of the incoming photon. For this reason, measurement of
the spectrum of the inelastically scattered outgoing photons is a
sensitive probe of the electron and phonon degrees of freedom in the
sample.

Therefore to describe theoretically the Raman spectrum of the rotated
double-layer graphene, we need to know its electron and phonon band
structures. Furthermore, we need to evaluate the matrix element of the
interaction between the electrons and light, and of the interaction
between the electrons and phonons. In the remainder of this section we
describe how we computed all of these quantities in the case of the
rotated double-layer graphene.  For simplicity, we only compute the
Raman intensity in the rotated double-layer graphene relative to the
Raman intensity in the single-layer graphene. Therefore, in our
calculation we don't include explicitly numerical prefactors common to
these two cases.

\subsection{Rotated double-layer graphene unit cell}
\label{sec:cell}

We start by defining the geometry of the rotated double-layer graphene
unit cell. The single-layer graphene unit cell consists of two carbon
atoms (A and B) arranged in a two-dimensional honeycomb lattice. The
rotated double-layer graphene consists of a stack of two single
graphene layers rotated with respect to each other by some angle
$\ang$. We define the angle $\ang$ as follows. We start from two
identical copies of single-layer graphene, translated along the axis
perpendicular to the graphene plane. Next, we perform the rotation of
one of the layers by angle $\ang$ around the axis passing through one
of the carbon atoms. We refer to these two layers either as the top
and the bottom layer, or as $L=1$ and $L=2$ layer.

\begin{figure}
\centering\includegraphics{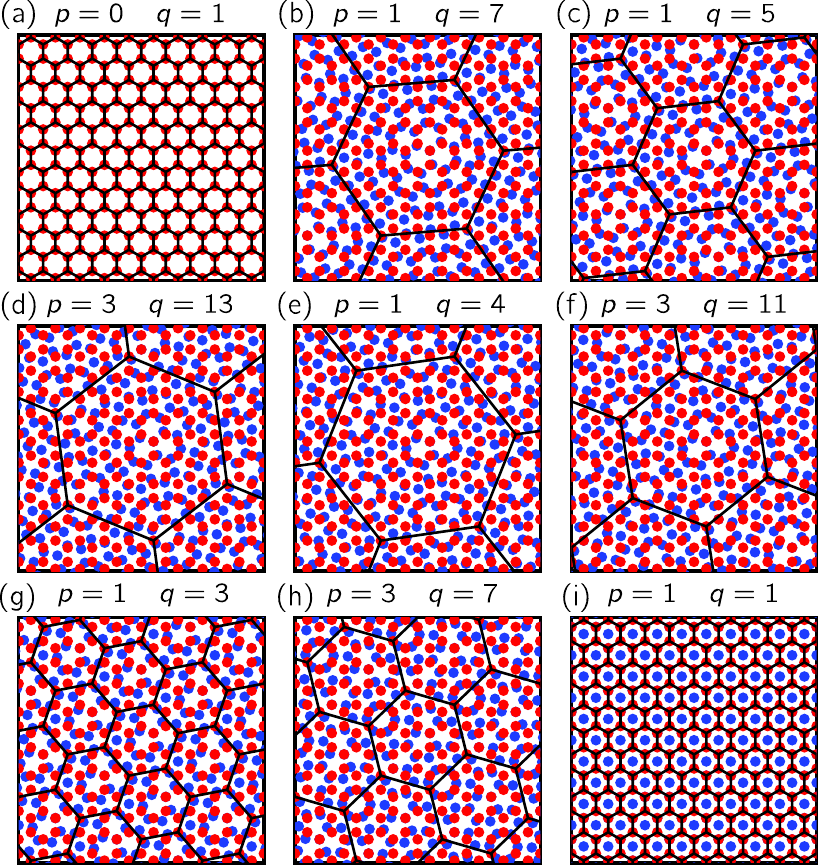}
\caption{(Color online.) A few examples of the rotated double-layer
  graphene for the special values of angle $\ang$ for which the
  structure is super-periodic. Carbon atoms from the fixed graphene
  layer are shown with the red dots. Blue dots indicate the carbon
  atoms in the layer rotated by the angle $\ang$. The unit cell
  (super-cell) of the rotated double-layer graphene is indicated with
  black lines.  Integers $p$ and $q$ for the each case are indicated
  on top of the each panel. Panel (a) corresponds to the double-layer
  graphene in which layers are not misaligned with respect to each
  other, $\ang=0^{\circ}$ (AA stacking). On the other hand, panel (i)
  corresponds to the case in which the angle $\ang$ is maximal,
  $\ang=60^{\circ}$ (AB stacking). Panels (b) through (h) cover range
  of angles from $0^{\circ}$ to $30^{\circ}$, specifically they are,
  $9.43^{\circ}$~(b), $13.17^{\circ}$~(c), $15.18^{\circ}$~(d),
  $16.43^{\circ}$~(e), $17.90^{\circ}$~(f), $21.79^{\circ}$~(g), and
  $27.80^{\circ}$~(h).}
\label{fig:unit_cell}
\end{figure}

In the case of our super-cell tight-binding method we work with angles
$\ang$ for which the resulting double-layer structure is periodic. As
shown in Ref.~\onlinecite{pap02} every periodic double-layer structure
is characterized by a pair of integers $p$ and $q$ up to a relative
translation of layers. Furthermore, angle $\ang$ is related to these
integers as
\begin{align}
\ang = \cos^{-1} \left( \frac{3q^2 - p^2}{3q^2+p^2} \right).
\end{align}
A few examples of the periodic double-layer structures are shown in
Fig.~\ref{fig:unit_cell}, and they all have characteristic moire
pattern resulting from the misalignment of two periodic structures.
The primitive unit cell of the rotated double-layer graphene is
indicated by black hexagons in Fig.~\ref{fig:unit_cell}. As can be
seen from the Fig.~\ref{fig:unit_cell}, the primitive cell of the
rotated double-layer graphene has a much larger area than the
single-layer primitive cell containing only two carbon atoms. As shown
in Ref.~\onlinecite{pap02}, area of the unit cell of the rotated
double-layer graphene is $N$ times larger than the single-layer unit
cell area, where $N$ is given as
\begin{align}
  N = \frac{{\rm gcd} (p,3)}{\left[ {\rm gcd}(p+3q,p-3q) \right]^2}
  \left( 3q^2 + p^2 \right).
  \label{eq:N}
\end{align}
Here ${\rm gcd}(a,b)$ is greatest common divisor of integers $a$ and
$b$. The smallest possible values of $N$ compatible with
Eq.~\ref{eq:N} are $7,13,19,31,37,43$, and $49$.

An $N$-fold increase of the real space primitive cell is accompanied
with a folding of the electron and phonon band structure in the
reciprocal space. The folded Brillouin zone area is smaller by a
factor of $1/N$ as compared to the single-layer Brillouin zone. The
thick black hexagon in Fig.~\ref{fig:bz} shows the Brillouin zone of
the rotated double-layer graphene for two choices of $p$ and $q$,
while the thick red and blue hexagons indicate the single-layer
Brillouin zones of the two individual layers. As one can see from the
figure, the rotated double-layer graphene Brillouin zone in these two
cases needs to be repeated six [for case from Fig.~\ref{fig:bz}(a)] or
twelve [Fig.~\ref{fig:bz}(b)] additional times in order to cover the
same area as the underlying single-layer graphene Brillouin zone.
Corresponding real-space super-cells for these two Brillouin zones are
shown in panels (g) and (h) of Fig.~\ref{fig:unit_cell}.

Throughout this paper we use a convention in which the wavevector from
the Brillouin zone of the rotated double-layer graphene is denoted by
$\k$. The reciprocal vector of the rotated double-layer graphene is
denoted by $\G$. The wavevector for the Brillouin zone of the two
single-layer graphene sheets will be denoted by $\kp$ and $\kpp$, for
the two sheets respectively. Corresponding reciprocal vectors of the
single-layer graphene sheets we will denote as $\Gp$ and $\Gpp$. We
always assume that vectors $\k$, $\kp$, $\kpp$, $\G$, $\Gp$, and
$\Gpp$ are given in the same coordinate system.

\subsection{Electrons}
\label{sec:elec}

We describe the electron wavefunctions in the rotated double-layer
graphene using a tight-binding model taking into account interaction
between the graphene layers. Low-energy electronic excitations in
graphene are well described by the carbon $\pi$-bonds. For this reason
our model includes only one $p_z$ orbital per carbon atom which we
will denote simply by $\phi(\r)$, for the carbon atom at the origin.
We further assume that the $\phi(\r)$ orbitals at the two neighboring
atomic sites are orthogonal to each other

Using orbitals $\phi(\r)$ we construct the Bloch-like tight-binding
basis functions $\chi_{\k j}(\r)$ for each $\k$-vector in the
Brillouin zone and for each site $j$ in the rotated double-layer
graphene unit cell (super-cell) as
\begin{align}
  \chi_{\k j}(\r) = \dsum_{\R} e^{i \k \cdot
    (\R + \t_j)} \phi (\r - \R - \t_j).
  \label{eq:d_basis}
\end{align}
Since rotated double-layer graphene consists of two graphene layers
and the primitive unit cell of each graphene layer has two carbon
atoms (A and B), the unit cell (super-cell) of the rotated
double-layer graphene has $4 N$ carbon atoms. Therefore index $j$
ranges from $1$ to $4N$. A sum is performed over all lattice vectors
$\R$, while the coordinate of $j$-th orbital in the primitive unit
cell is given by the vector $\t_j$.

The functions $\chi_{\k j}(\r)$ satisfy the periodicity requirement of
the Bloch theorem so we can write the $m$-th electron eigenstate
$\psi_{\k}^{m} (\r)$ simply as a linear combination of the basis
functions $\chi_{\k j}(\r)$,
\begin{align}
  \psi_{\k}^{m} (\r) = \dsum_j C_{\k j}^{m} \chi_{\k j}
  (\r).
  \label{eq:psi_states}
\end{align}
The band index $m$ again ranges from $1$ to $4N$, while only half
($2N$) of these bands are assumed to be occupied for undoped systems
(as in the single-layer graphene case).

Next, by solving the Schrodinger equation for the electrons in the
$\chi_{\k j} (\r)$ basis we obtain a set of $C_{\k j}^{m}$
coefficients at the each vector $\k$ of interest. In order to
construct the Schrodinger equation, we use the Slater-Koster
parametrization from Ref.~\onlinecite{pap03} fitted to the density
functional theory calculation of the rotated double-layer graphene. We
also rescale all tight-binding hopping parameters from
Ref.~\onlinecite{pap03} by $18 \%$ to account for the GW computed
self-energy effects\cite{pap32-1,pap32-2,pap32-3,pap32-4}.

\subsubsection{Unfolding of the electron states}
\label{sec:unfold}

We now describe a procedure in which one can rewrite (unfold) the
rotated double-layer graphene electron wavefunction in terms of the
single-layer graphene basis functions. This procedure will be useful
later in the computation of the electron-phonon matrix element of the
rotated double-layer graphene.

In a close relation to the Eq.~\ref{eq:d_basis} let us now define the
following basis functions
\begin{align}
  \xi_{\k \alpha}(\r) & =
  \dsum_{j \rightarrow \alpha} \dsum_{\R} e^{i \k \cdot (\R + \t_j)}
  \phi (\r - \R - \t_j).
  \label{eq:s_basis}
\end{align}
Here, $\alpha = (L,\lambda)$ is a composite index where $L=1, 2$
indexes the graphene layers and $\lambda=A, B$ indexes the A and B
carbon atoms. The sum over index $j$ in Eq.~\ref{eq:s_basis} is
performed over all atoms of type $\alpha$, i.e. over all A or B carbon
atoms in either first or second graphene layer.

Functions $\xi_{\k \alpha}(\r)$ respect the periodicity of the
single-layer graphene sheet in the same way that $\chi_{\k j} (\r)$
respect the periodicity of the rotated double-layer graphene. For this
reason if we consider indices $\alpha$ from either first or second
layer ($L=1$ or $2$), functions $\xi_{\k \alpha}(\r)$ become
Bloch-like tight-binding basis functions of the first or the second
single-layer graphene [in the same way in which $\chi_{\k j} (\r)$ are
the basis functions of the rotated double-layer graphene].

Computing the overlap between the rotated double-layer graphene
wavefunction $\psi_{\k}^{m} (\r)$ and the single-layer graphene basis
function $\xi_{\kap \alpha}(\r)$ gives
\begin{align}
  \dpr{\xi_{\kap \alpha}}{\psi_{\k}^{m}} = n_{\rm sc} \delta_{\kap, \k
    + \G} 
  P_{\k \G \alpha}^{m},
\label{eq:bas_over}
\end{align}
where we have defined quantity $P_{\k \G \alpha}^{m}$ as
\begin{align}
  P_{\k \G \alpha}^{m}=
  \dsum_{j \rightarrow \alpha} C_{\k j}^{m} e^{-i \G \cdot \t_j }.
  \label{eq:def_P}
\end{align}
In Eq.~\ref{eq:bas_over}, the term $\delta_{\kap, \k + \G}$ equals $1$
only if $\kap$ and $\k$ differ by one of the rotated double-layer
reciprocal vector $\G$, while $n_{\rm sc}$ is the total number of the
super-cells in the entire sample.  Since the $\ket{\xi_{\kap \alpha}}$
basis is complete, Eq.~\ref{eq:bas_over} implies that the electron
wavefunction $\ket{\psi_{\k}^{m}}$ of the rotated double-layer
graphene can be rewritten (unfolded) in terms of the basis functions
$\ket{\xi_{\k + \G \alpha}}$ of the {\it single-layer} graphene (here
$\G$ are the reciprocal vectors of the rotated double-layer graphene).
Furthermore, unfolding amplitude of the rotated double-layer graphene
electron wavefunction $\ket{\psi_{\k}^{m}}$ in terms of the
single-layer graphene basis function $\ket{\xi_{\k+\G \alpha}}$ is
given by quantity $P_{\k \G \alpha}^{m}$ defined in
Eq.~\ref{eq:def_P}.

\begin{figure}
\centering\includegraphics{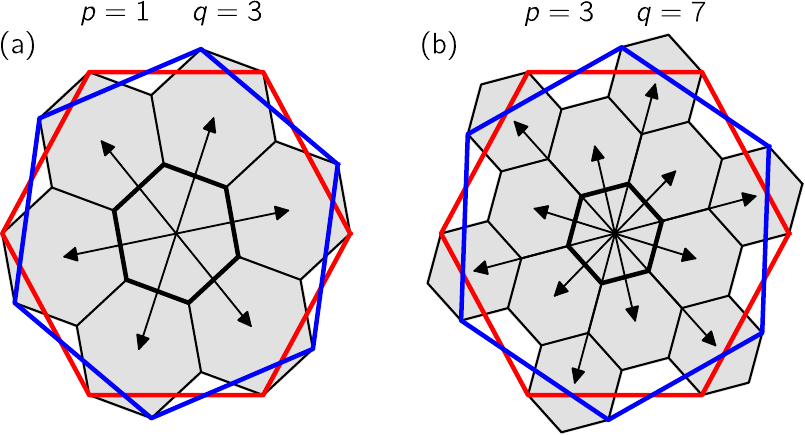}
\caption{(Color online.) Unfolding of the rotated double-layer
  Brillouin zone (thick black line) onto two single-layer Brillouin
  zones (red and blue). Panel (a) corresponds to $p=1$, $q=3$,
  $\ang=21.79^{\circ}$, $N=7$ [corresponding real-space cell is shown
  in panel (g) of Fig.~\ref{fig:unit_cell}], while panel (b)
  corresponds to $p=3$, $q=7$, $\ang=27.80^{\circ}$, $N=13$
  [corresponding real-space cell is shown in panel (h) of
  Fig.~\ref{fig:unit_cell}]. Arrows indicate $\G$ vectors from the set
  $\cal G$ (see main text). Determination of the set of vectors $\cal
  G$ is relatively easy for these choices of $p$ and $q$, while it
  becomes more involved for some other choices since some elements of
  $\cal G$ must point outside of the single-layer Brillouin zone.}
\label{fig:bz}
\end{figure}

We perform the unfolding procedure on a fixed set of $\G$ vectors,
${\cal G}$, such that the following two constraints are satisfied.
First, vectors $\k+\G$ with two different choices of $\G$ from the set
$\cal G$ never differ from each other either by $\Gp$ or $\Gpp$ (
reciprocal vectors of two single-layer graphene sheets), as this would
lead to double counting.  Second, every unique single-layer wavevector
$\kp$ or $\kpp$ can be written as $\k+\G$ for some $\G$ from $\cal G$
and $\k$ from the double-layer graphene Brillouin zone.  Black arrows
on Fig.~\ref{fig:bz} indicate two examples of a set of $\G$ vectors
$\cal G$ satisfying these constraints.

\subsection{Phonons}
\label{sec:phons}

In this work we assume that the interaction between the two layers of
graphene does not affect the phonon band structure of the rotated
double-layer graphene. Nevertheless, working in the Brillouin zone of
the rotated double-layer graphene we need to take into account that
the set of the single-layer phonons at wavevectors $\q+\G$ are folded
to the wavevector $\q$ for all $\G$ from ${\cal G}$. Unlike for the
electron states, which do get affected by the interaction between the
two graphene layers, here the unfolding procedure corresponds simply
to the relabeling of states. In the folded double-layer Brillouin zone
we denote phonon states with the wavevector label $\q$ and the branch
label $\nu$. On the other hand, in the unfolded single-layer Brillouin
zone, this same phonon would be labeled with wavevector $\q+\G$ with
vector $\G$ chosen from the set $\cal G$. Therefore labels $\nu$ and
$\G$ are interchangeable for a unique phonon branch (corresponding for
example either to the G or the 2D peak) of the single-layer graphene.

In our calculations of the Raman G peak we use the phonon frequency of
the G phonon to equal 1561~cm$^{-1}$, as found in
Ref.~\onlinecite{mohr2007}. The G phonon atomic displacement pattern
can uniquely be determined from the representation theory analysis of
the graphene space group.

Calculation of the Raman 2D peak is more complex than that of the G
peak, since it involves emission of two phonons with arbitrary (and
opposite) momentum. For this reason we need information about the 2D
phonon frequencies in a relatively large region of the phonon
Brillouin zone close to the K-point (phonons far away from the K-point
give negligible contribution to the Raman 2D peak). It was shown in
Ref.~\onlinecite{pap34} that the Raman 2D peak profile relies strongly
on the compensation between the phonon and the electron trigonal
warpings, which are shown to be of a different sign.  Therefore, the
Raman 2D spectrum in graphene is very sensitive to the details of the
phonon band structure. For this reason we use as an input to our
calculations a high-order polynomial fit to the computed 2D phonon
frequencies from Ref.~\onlinecite{pap34}, and we include the computed
trigonal warping effect. Atomic displacements of 2D phonons are
inferred only at the K (and K') points of the Brillouin zone by the
representation theory analysis. The atomic displacement pattern of the
phonons near the K (or K') point are approximated by the displacement
pattern at K (or K') point.

\subsection{Electron-light interaction}
\label{sec:elec_light}

In the weak field approximation\footnote{In this work we neglect
  $\sim{\bf A}^2$ term in the electron-light interaction Hamiltonian.
  We discuss validity of this approximation in the context of Raman G
  peak calculation in Sec.~\ref{sec:G}.}  interaction of electrons
with light is given by the following operator
\begin{equation}
H^{\rm light} = \frac{i e}{m} {\bf A} \cdot \bm{\nabla}
\label{eq:e-light}
\end{equation}
where ${\bf A}$ is the vector potential of the electromagnetic field
of light in the Coulomb gauge ($\hbar=1$). Therefore the matrix
element of this operator between two electron states $\bra{f}$ and
$\ket{i}$ up to a constant equals $\P\cdot\me{f}{\bm{\nabla}}{i}$,
where $\P$ is the polarization direction of the incoming or outgoing
light. Expressing electron states $\bra{f}$ and $\ket{i}$ in terms of
the basis functions $\phi (\r)$ we are left with computing
$\me{\phi}{\bm{\nabla}}{\phi'}$ where $\bra{\phi}$ and $\ket{\phi'}$
are the tight-binding basis orbitals $\phi (\r)$ at the different
atomic sites. We work under approximation\cite{gruneis2003} that the
matrix element $\me{\phi}{\bm{\nabla}}{\phi'}$ is exactly zero between
$\bra{\phi}$ and $\ket{\phi'}$ that are not on the first-neighbor
sites in the same graphene layer. Additionally, assuming $p_{\rm
  z}$-like symmetry of the $\phi (\r)$ orbitals one can easily show
that under this approximation all matrix elements
$\me{\phi}{\bm{\nabla}}{\phi'}$ can be determined up to a single
constant prefactor.

\subsection{Electron-phonon interaction}
\label{sec:elec_phonon}

Interaction between electrons and phonons $H^{\rm ph}_{{\bf q} \nu}$
is described in terms of the deformation potential $\delta V_{{\bf q}
  \nu}$. The deformation potential is defined as a change in the
effective potential experienced by electrons as a result of the phonon
excitation from the $\nu$-th phonon branch with a wavevector $\bf q$.
Similarly as in the case of an electron-light interaction, using
symmetry and taking into account the interaction between the nearest
neighbors, the electron-phonon matrix elements can be computed up to a
constant prefactor for both G and 2D phonon modes.

Since we assumed no changes to the phonon band structure coming from
the interaction between the graphene layers, we compute the
electron-phonon matrix element using the same electron-phonon
interaction operator as in the case of a single-layer graphene. We
start from the electron-phonon matrix element between any two rotated
double-layer graphene states $\bra{\k}_{{\rm dl}}$ and $\ket{\k + {\bm
    q}}_{{\rm dl}}$ (dropping electron band indices),
\begin{align}
\bra{\k }_{{\rm dl}} H_{ {\bm q} \nu }^{\rm ph} 
\ket{\k+{\bm q}}_{{\rm dl}}.
\end{align}
Next we express the rotated double-layer states in the basis of the
single-layer states using Eqs.~\ref{eq:bas_over} and \ref{eq:def_P}
(we absorb coefficients $P_{\k \G \alpha}^{m}$ into $\bra{\dots}_{{\rm
    sl}}$ and $\ket{\dots}_{{\rm sl}}$ for simplicity),
\begin{align}
  \left[ \dsum_{\G_1 \in {\cal G}} \bra{\k +\G_1}_{{\rm sl}} \right]
  H_{ {\bm q} \nu }^{\rm ph} \left[ \dsum_{\G_2 \in {\cal G}}
    \ket{\k+{\bm q}+\G_2}_{{\rm sl}} \right].
\end{align}
Furthermore, here we also drop sum over $\alpha$ for simplicity.
Remembering that the branch index $\nu$ for the folded phonon band
structure is just a relabeling of vectors $\G_3$ from the set $\cal G$
we can write the electron-phonon matrix element as
\begin{align}
  \dsum_{\G_1 \in {\cal G}}
  \dsum_{\G_2 \in {\cal G}}
   \bra{\k +\G_1}_{{\rm sl}}
  H_{ {\bm q} + \G_3 }^{\rm ph} 
  \ket{\k+{\bm q}+\G_2}_{{\rm sl}}.
\end{align}
Next, as we mentioned earlier, we assume that the electron-phonon
interaction {\it operator} in the rotated double-layer graphene is the
same as in the single-layer graphene. For this reason, the
electron-phonon matrix operator conserves the crystal momentum of the
single-layer graphene. Therefore, only one of the $\G_3$ vectors will
give a non-zero contribution to the electron-phonon matrix element,
\begin{align}
  \dsum_{\G_1 \in {\cal G}} 
  \dsum_{\G_2 \in {\cal G}}
  \bra{\k +\G_1}_{{\rm sl}} 
  H_{ {\bm q} + \G_2 - \G_1 }^{\rm ph} 
  \ket{\k+{\bm q}+\G_2}_{{\rm sl}}.
  \label{eq:phonG1G2}
\end{align}

\subsection{Raman intensity}
\label{sec:raman_int}

Using standard perturbation technique methods\cite{pap18,pap34},
one can show that the intensity of the outgoing photon at frequency
$\omega_{\rm out}$ for the first-order Raman process can be computed
as
\begin{align}
  I_1 (\omega_{\rm out}) \sim \sum_{\nu} \left| \sum_{{\rm A} {\rm
        B}} K_{A B}^{\nu} \right|^2 \delta(\omega_{\rm in} -
  \omega_{{\bm 0}}^{\nu} - \omega_{\rm out}),
  \label{eq:firstRaman}
\end{align}
while for the second-order Raman process it is given by
\begin{align}
  I_2 (\omega_{\rm out}) \sim \sum_{\q \nu \mu} \left| \sum_{{\rm A} {\rm B}
      {\rm C}} K_{A B C}^{\q \nu \mu} \right|^2 \delta(\omega_{\rm
    in} - \omega_{-\q}^{\nu} - \omega_{\q}^{\mu} - \omega_{\rm out}).
  \label{eq:secondRaman}
\end{align}
Here frequency of the $\nu$-th ($\mu$-th) phonon branch with the
momentum $\q$ is denoted as $\omega_{\q}^{\nu}$ ($\omega_{\q}^{\mu}$),
while the incoming light frequency is denoted as $\omega_{\rm in}$.
Furthermore, here for simplicity we always assume that the
phonon-dependent terms (phonon frequencies, electron-phonon matrix
elements) appearing in the first-order Raman process are due to the G
mode, while those appearing in the second-order Raman process are due
to the 2D mode. Scattering amplitudes $K_{A B}^{\nu}$ and $K_{A B
  C}^{\q \nu \mu}$ are summed over all virtually excited states $A$,
$B$, and $C$. Sum in Eqs.~\ref{eq:firstRaman} and \ref{eq:secondRaman}
is performed coherently over the electron states and incoherently over
the phonon states. Delta functions ensure the conservation of energy.

In both Eqs.~\ref{eq:firstRaman} and \ref{eq:secondRaman} we focus
only on the processes involving emission, not absorption, of phonons,
and we work at zero temperature. Furthermore, we are neglecting the
momentum of the photon. Therefore to conserve total momentum, the
emitted phonon in the first-order Raman process must have zero
momentum. In the second-order process momentum of one phonon ($\q$)
must be compensated by that of the other phonon ($-\q$).  For this
reason, the first sum in Eq.~\ref{eq:firstRaman} is performed over
zero-momentum phonons, from arbitrary phonon branch $\nu$.  Similarly,
first sum in the Eq.~\ref{eq:secondRaman} is performed over all pairs
of phonons with momenta $\q$ and $-\q$, from possibly different phonon
branches $\nu$ and $\mu$.

\begin{figure}
\centering\includegraphics{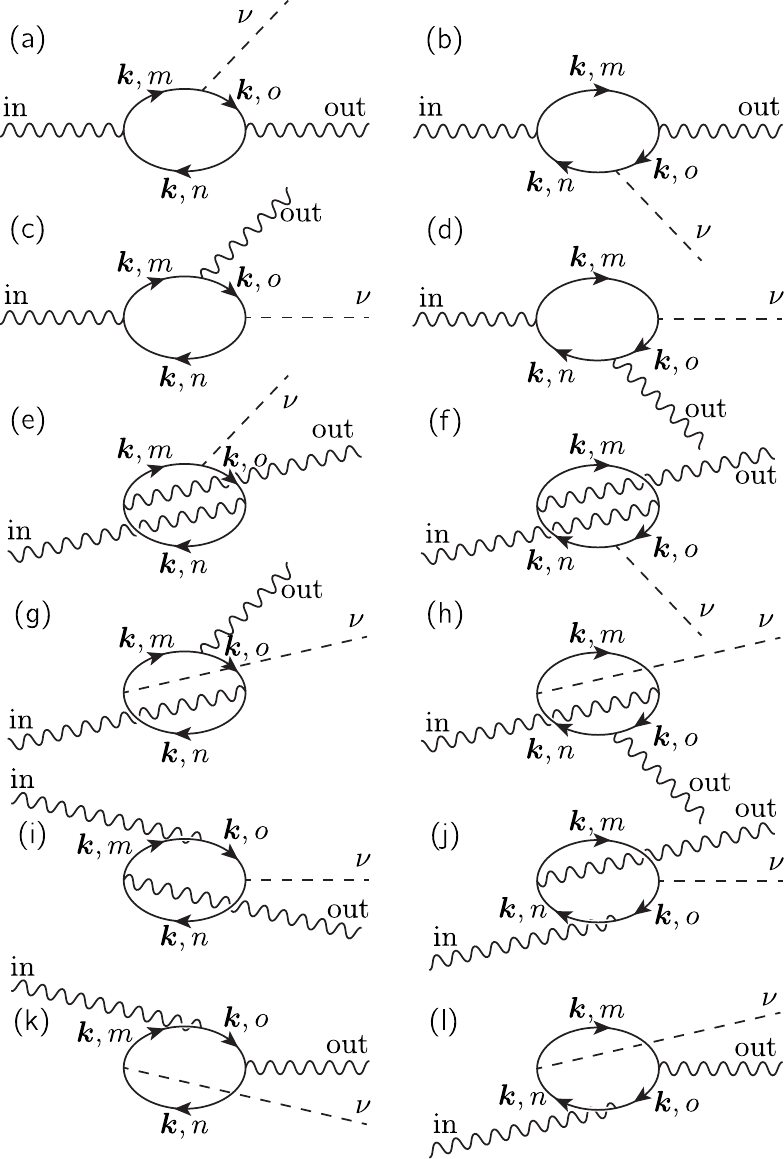}
\caption{Feynman diagrams included in the first order Raman
  calculation for the Raman G peak. Time is increasing from the left
  to the right, photons are indicated with wavy line while phonon is
  shown with dashed line. Electrons and holes are drawn with arrows in
  the opposite direction with respect to time. Explicit expression for
  the diagram in the panel (a) is given in Eq.~\ref{eq:example_G}.}
\label{fig:diag_G}
\end{figure}

\begin{figure}
\centering\includegraphics{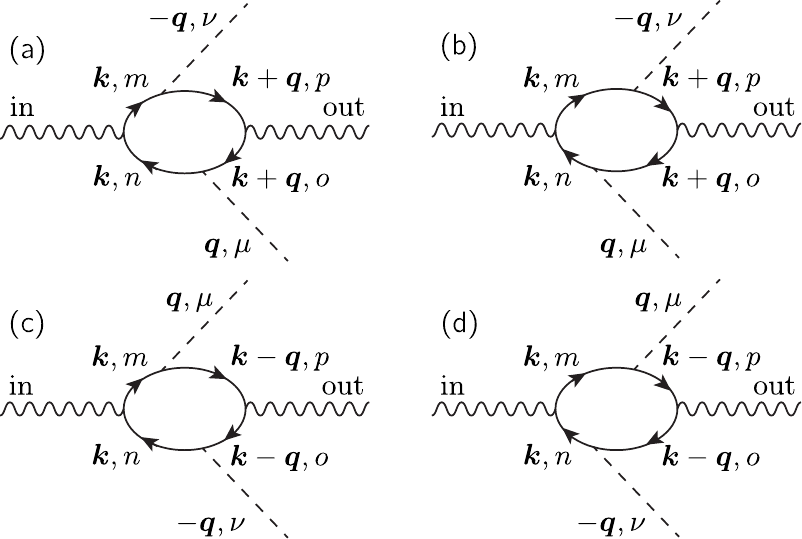}
\caption{Feynman diagrams included in the second order Raman
  calculation for the Raman G peak. Conventions are as in
  Fig.~\ref{fig:diag_G}.  Explicit expression for the diagram in
  panel (a) is given in Eq.~\ref{eq:example_2D}.}
\label{fig:diag_2D}
\end{figure}

Scattering amplitudes $K_{A B}^{\nu}$ and $K_{A B C}^{\q \nu \mu}$ are
most easily represented graphically using Feynman diagrams as in
Figs.~\ref{fig:diag_G} and \ref{fig:diag_2D}. Explicit expressions for
these diagrams can be found in Refs.~\onlinecite{pap18} and
\onlinecite{pap34}, here we provide as an example contribution from
Fig.~\ref{fig:diag_G}(a) for the first order Raman process,
\begin{align}
  K_{\k m n o}^{\nu}=
  &
  \me{\k n        }{H^{\rm light}_{\rm out} }{\k \overset{\star}{o}}
  \me{\k \overset{\star}{o}}{H_{{\bm 0} \nu}^{\rm ph}}{\k \overset{\star}{m}}
  \me{\k \overset{\star}{m}}{H^{\rm light}_{\rm in}  }{\k n        }
  \cdot
  \notag \\
  &\left(
    \omega_{\rm in}
    - \omega_{ {\bm 0}}^{\nu}
    - \epsilon_{\k}^{o}
    + \epsilon_{\k}^{n}
    - i \frac{\gamma}{2}
  \right)^{-1}
  \cdot
  \notag \\
  & \left(
    \omega_{\rm in}
    - \epsilon_{\k}^{m}
    + \epsilon_{\k}^{n}
    - i \frac{\gamma}{2}
  \right)^{-1}.
\label{eq:example_G}
\end{align}
Similarly, we also provide an explicit expression for the contribution of
the second order Raman process from Fig.~\ref{fig:diag_2D}(a),
\begin{align}
  K_{\k m n o p}^{\q \nu \mu}=
  &
\me{\k+{\bm q} o}{H^{\rm light}_{\rm out}  }{\k+{\bm q} \overset{\star}{p}}
\me{\k         n}{H_{ {\bm q} \mu}^{\rm ph}}{\k+{\bm q} o        } \cdot 
\notag \\
&
\me{\k+{\bm q} \overset{\star}{p}}{H_{-{\bm q} \nu}^{\rm ph}}{\k \overset{\star}{m}}
\me{\k         \overset{\star}{m}}{H^{\rm light}_{\rm in}   }{\k n        } \cdot
\notag \\
&\left(
  \omega_{\rm in}
- \omega_{-{\bm q}}^{\nu}
- \omega_{ {\bm q}}^{\mu}
- \epsilon_{\k + {\bm q}}^{p}
+ \epsilon_{\k + {\bm q}}^{o}
- i \frac{\gamma}{2}
\right)^{-1}
\cdot
\notag \\
&\left(
  \omega_{\rm in}
- \omega_{-{\bm q}}^{\nu}
- \epsilon_{\k + {\bm q}}^{p}
+ \epsilon_{\k}^{n}
- i \frac{\gamma}{2}
\right)^{-1} \cdot
\notag \\
& \left(
  \omega_{\rm in}
- \epsilon_{\k}^{m}
+ \epsilon_{\k}^{n}
- i \frac{\gamma}{2}
\right)^{-1}.
\label{eq:example_2D}
\end{align}
Electron bands indices in Eqs.~\ref{eq:example_G} and
\ref{eq:example_2D} are $m,n,o,p$, while the phonon branch indices are
$\nu$ and $\mu$. Electron eigenenergy at wavevector $\k$ and for the
band $m$ is indicated with $\epsilon_{\k}^{m}$. Sum of the electron
and the hole linewidth is given by $\gamma$, which we discuss in more
detail in Sec.~\ref{sec:params}. Empty electron states in
Eqs.~\ref{eq:example_G} and \ref{eq:example_2D} have the symbol
$^{\star}$ over their band indices.

We find that it is important to include all of the
first-order diagrams for the Raman G peak (as shown in
Fig.~\ref{fig:diag_G}). For the 2D peak we include only diagrams shown
in Fig.~\ref{fig:diag_2D}, since other permutations give much smaller
contributions in the case of a single-layer graphene, see
Ref.~\onlinecite{pap34} for more details.

\subsection{Remaining parameters}
\label{sec:params}

Here we discuss remaining parameters and calculation details used in
this work. For the electron and hole linewidth $\gamma$ appearing in
Eqs.~\ref{eq:example_G} and \ref{eq:example_2D} for the Raman G and 2D
peak intensity we use $\frac{\gamma}{2}=190$~meV and
$\frac{\gamma}{2}=201$~meV for the 1.96~eV and 2.41~eV photon energy
calculation respectively, independent of electron wavevector $\k$. We
choose this value of electron and hole linewidth in order to reproduce
the Raman G peak enhancement factor (discussed later, in
Sec.~\ref{sec:G}) consistent with experiment done with 1.96~eV
incoming photon energy\cite{kim2012}.  Nevertheless, we find this
value to be somewhat consistent with the sum of linewidths coming from
the electron-phonon\cite{pap34} (32 and 43~meV for 1.96~eV and 2.41~eV
photon energy calculation respectively) and electron-electron
interaction ($\sim 100$~meV\cite{jornada}).  Using the electron
linewidth coming just from the electron-phonon interaction (as done in
Ref.~\onlinecite{pap34}) would have resulted in a much larger Raman G
peak enhancement. The linewidth $\frac{\gamma}{2}$ used for the
2.41~eV incoming photon energy calculation was computed from the value
for the 1.96~eV incoming photon energy by including the difference in
the estimated electron-phonon linewidths ($43$~meV$-32$~meV=$11$~meV).

To speed up the convergence of the Raman calculation in the case of a
rotated double-layer graphene we interpolate various electron and
phonon quantities from a coarser reciprocal space grid onto a finer
grid. Furthermore, we neglect Raman amplitudes for which
unfolding intensity, electron-light matrix element, or
electron-phonon matrix element fall below a certain threshold value.
We check that our results are fully converged with respect to
this threshold.

\subsection{Continuum model method}
\label{sec:cont}

In this work, we also make use of the continuum model developed in
Ref.~\onlinecite{Bistritzer2011}, in order to confirm our super-cell
tight-binding based calculation of the Raman 2D peak. The continuum
model, as compared to the super-cell tight-binding method, uses an
electron wavefunction in the restricted Hilbert space. The Brillouin
zone folding in the super-cell tight-binding calculation implies that
the interaction between the graphene layers introduces hybridization
between states at $N$ different wavevectors (state with wavevectors
$\k$ is hybridized with states $\k+\G$, $\G \in \cal G$).  On the
other hand, in our continuum model, we select only a subset of vectors
$\G$ from $\cal G$ for which the interlayer hybridization is the
strongest.  In particular, a state with wavelength $\kp$ in layer
$L=1$ hybridizes only with the electron states in layer $L=2$ with
wavelength $\kp+\Gp$. In the case of our continuum model calculation,
we consider three reciprocal vectors $\Gp$ of layer $L=1$ for which
the norm $|\kp+\Gp|$ is minimized. This approximation can be justified
with a perturbative calculation, as in
Ref.~\onlinecite{Bistritzer2011}.

Furthermore, as compared to the super-cell tight-binding calculation,
in our continuum model calculation we are using a simple
parametrization\cite{Bistritzer2011} of the interaction strength
between the graphene layers that depends on only one parameter, $c$.
Additionally, in our continuum model we neglect trigonal warping of
the single-layer graphene band structure, and assume perfectly linear
Dirac cone band structure parametrized with the band velocity $v_{\rm
  F}$.  Following Ref.~\onlinecite{Bistritzer2011}, we take
$v_F=10^6$~${\rm m/s}$ and $c=0.11$~eV. Since we neglect the trigonal
warping effect, Raman G peak intensity vanishes in the continuum
model.  This is because the electron-phonon matrix elements change
sign under the operation ${\bf K}+{\bm \kappa} \rightarrow {\bf
  K}-{\bm \kappa}$ in the continuum model,\cite{Ando2006} where ${\bf
  K}$ is the Dirac point of single-layer graphene.  The product of the
electron-light matrix elements and the energy denominators in
Eq.~\ref{eq:example_G} do not change under this operation in the
continuum model. The sum over scattering amplitudes in
Eq.~\ref{eq:firstRaman} therefore vanishes in the continuum model
without trigonal warping.  For this reason we use continuum model only
to compute the Raman 2D peak.

The electron-light matrix element, the electron-phonon matrix element,
and the Raman intensity in the continuum model are computed as in the
super-cell based method. We use the same value of the electron
linewidth in the two calculations.

\section{Results and discussion}
\label{sec:results}

In this section we present results of our calculations of the Raman
spectra in the rotated double-layer graphene.

\subsection{Electronic structure}
\label{sec:el}

\begin{figure}
\centering\includegraphics{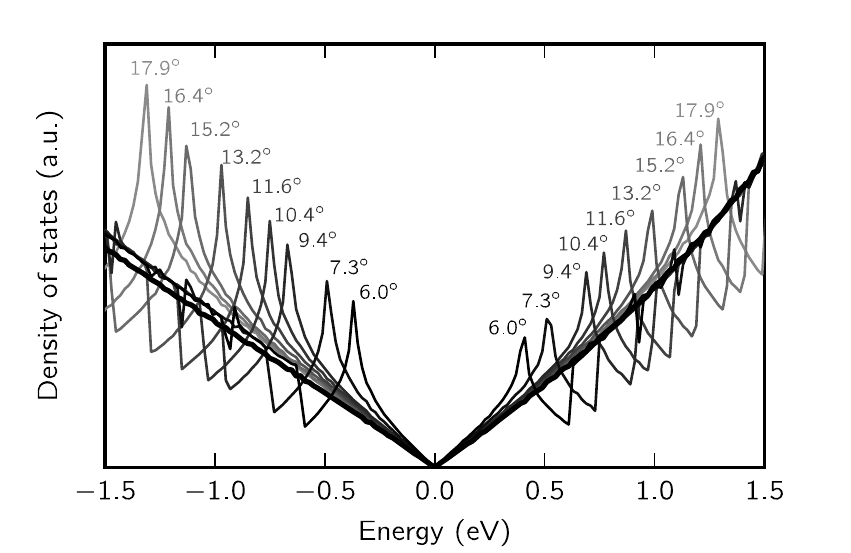}
\caption{(Color online.) Density of states in our super-cell
  tight-binding model near the Fermi level (at the zero energy) for
  the rotated double-layer graphene at varying angles $\ang$ from
  $6.0^{\circ}$ to $17.9^{\circ}$. Lighter gray lines correspond to
  larger values of $\ang$. For each $\ang$ we find two large Van Hove
  singularities next to each other, with similar energy.  Step-like
  singularity arises from the energy maximum or minimum (as a function
  of momentum) while the logarithmic divergence arises from the energy
  saddle point. As angle $\ang$ is increased, these singularities move
  further away from the Fermi level (compare lighter and darker gray
  lines in the figure).  Thick black line is showing the density of
  states of a single-layer graphene, multiplied by two, so that it can
  be compared more easily to the rotated double-layer graphene case.}
\label{fig:dos}
\end{figure}

We start with a discussion of the electronic structure of the rotated
double-layer graphene. Density of states for varying angles
$\ang$ are given in Fig.~\ref{fig:dos} with thin gray lines,
while that for the single-layer graphene is given with a thick black
line. The density of states of the single-layer graphene in this range
of energies linearly increases with the energy as one moves away from
the Fermi level (Fermi level is at the zero energy in
Fig.~\ref{fig:dos}).  This linear dependence of the density of states
originates from the well known Dirac cones at the Brillouin zone
corners of the single-layer graphene band structure.

The two graphene layers in the rotated double-layer graphene are
rotated with respect to each other by an angle $\ang$. For this
reason, the Dirac cones from each layer are not exactly on top of each
other (in the reciprocal space) but are instead rotated with respect
of each other by the angle $\ang$. Therefore, the two Dirac cones are
overlapping only in a small region of the reciprocal space, and
position of this overlap in the reciprocal space depends on the angle
$\ang$. In this overlap region interaction between the two layers
opens a hybridization gap, which in turn leads to the occurrence of
prominent Van Hove singularities both in the occupied and empty
states, whose position again depends on $\ang$. For example in
$\ang=6.01^{\circ}$ case Van Hove singularities occur near
$\pm0.5$~eV, while for the $\ang=13.17^{\circ}$ case Van Hove
singularities occur near $\pm1.0$~eV.

\subsection{Raman G peak}
\label{sec:G}

As shown in the Fig.~\ref{fig:dos}, the energy at which the Van Hove
singularities occur in the rotated double-layer graphene depends on
the angle $\ang$. For larger values of $\ang$, the Van Hove
singularities occur further away from the Fermi level. In particular,
for the larger value of angle $\ang$ the Van Hove singularity of the
occupied states are moved to lower energies, while those of the empty
states are moved to the larger energies. When separation between the
Van Hove singularities of the empty and occupied states matches the
incoming photon energy, we expect to see changes of the rotated
double-layer graphene Raman spectrum. Angle $\ang$ for which the
incoming photon energy is close to the separation between the Van Hove
singularities we will refer to as the {\it critical angle}.

\begin{figure}
\centering\includegraphics{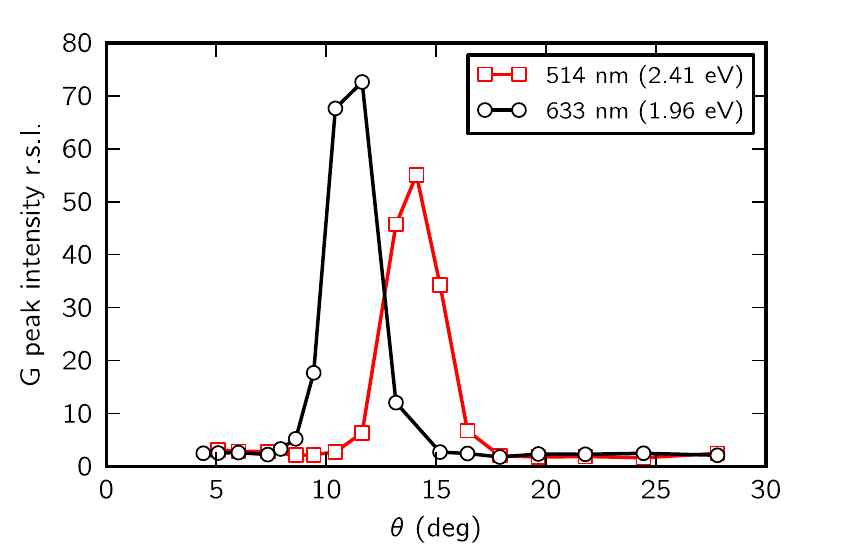}
\caption{(Color online.) Calculated Raman G peak intensity as a
  function of angle $\ang$ for two incoming photon energies [1.96~eV
  in black and 2.41~eV in red (gray)]. The range of angle $\ang$ shown
  is from $0^{\circ}$ to $30^{\circ}$. For range
  $30^{\circ}<\ang<60^{\circ}$ we find almost the same Raman G peak
  intensity for $\ang=30^{\circ}+\Delta$ as for the
  $\ang=30^{\circ}-\Delta$ case.  Intensity is measured relative to a
  single-layer graphene. We find $\sim 70$ fold enhancement in the
  Raman G peak intensity for 1.96~eV incoming photon energy near the
  critical angle, $10^{\circ}$. This enhancement shifts to the higher
  angles $\ang$ for higher incoming photon energy (2.41~eV),
  consistent with shift in the Van Hove singularity. At angles away
  from both sides of the critical angle, we find Raman G peak
  enhancement close to 2 (as would be expected in the limit of no
  interaction between the layers).  Comparison with experimental data
  (in good agreement with our calculation) is shown in
  Ref.~\onlinecite{kim2012}.}
\label{fig:G}
\end{figure}

The computed Raman G peak intensity in the rotated double-layer
graphene is given in Fig.~\ref{fig:G} as a function of angle $\ang$
for two different incoming photon energies (black and red line). The
Raman G peak intensity in Fig.~\ref{fig:G} is given in terms of the
intensity of a single-layer graphene. We find a $\sim 70$ fold
enhancement of the Raman G peak intensity at angles $\ang$ close to
the critical angle, $10^{\circ}$ (1.96~eV incoming photon energy,
black line in Fig.~\ref{fig:G}). At angles below and above this
critical angle we find that the Raman G peak enhancement factor is
close to 2.  Therefore, in that region of angles $\ang$ Raman signal
of the rotated double-layer graphene is almost the same as that of two
independent graphene sheets.  Furthermore, we also find that the G
peak enhancement shifts to the higher angles $\ang$ with higher
incoming photon energy (red line in Fig.~\ref{fig:G}). This behavior
we attribute to the shift in the energy of the Van Hove singularity as
a function of angle $\ang$, as observed already in Fig.~\ref{fig:dos}.

Unlike the Raman 2D peak, the Raman G peak in graphene is a single phonon
process and therefore its width and peak position depend solely on the
phonon lifetime and frequency. We assumed in our calculation that the
phonon lifetime and frequency are not affected by the interaction
between the two graphene layers. For this reason, Raman G peak width
and position are independent of the angle $\ang$, in agreement with
experimental observations in Ref.~\onlinecite{kim2012}.

\begin{figure}
\centering\includegraphics{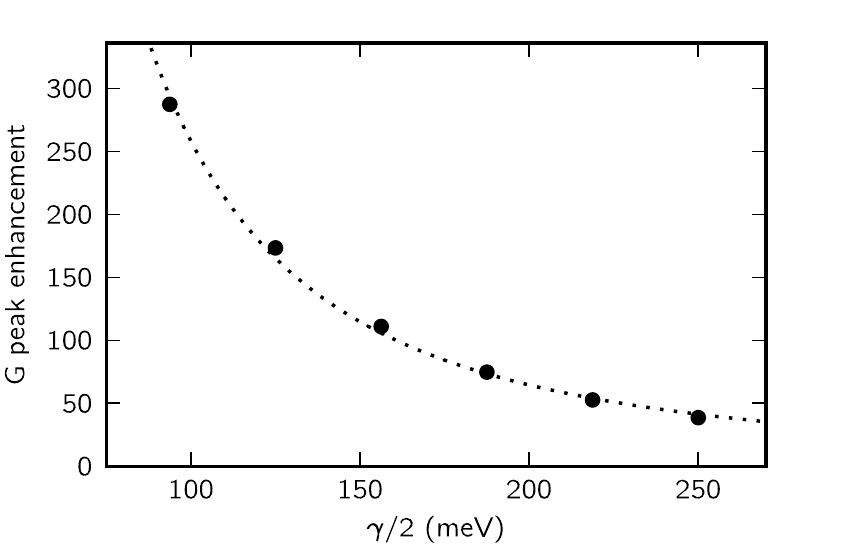}
\caption{Dependence of the Raman G peak enhancement (relative to a
  single-layer) at the critical angle on the electron and the hole
  lifetime $\gamma$. The incoming photon energy in this calculation
  equals 1.96~eV. Fitted functional dependence of the Raman G peak
  enhancement ($G_{\rm enh}$) is indicated with a dotted line, and
  equals $G_{\rm enh} = 2.58 ~(\mathrm{eV}^{2}) (\gamma/2)^{-2}$.}
\label{fig:ellf}
\end{figure}

We find a very strong dependence of the Raman G peak enhancement at
the critical angle on the effective electron and hole linewidth
$\gamma$ appearing in Eq.~\ref{eq:example_G}. Dependence of the Raman
G peak enhancement at the critical angle on the value of parameter
$\gamma$ is shown in Fig.~\ref{fig:ellf}. Dotted line in
Fig.~\ref{fig:ellf} is a fit to the function $\sim \gamma^{-2}$. As
already mentioned in Sec.~\ref{sec:params} due to this strong
dependence of the Raman G peak enhancement on $\gamma$, we have chosen
value of $\gamma$ which gives Raman G peak enhancement in agreement
with experiment at 1.96~eV incoming photon energy.  Nevertheless, the
value of $\gamma$ we obtained is consistent with that obtained from
the electron-phonon and electron-electron interaction estimates.

In the Raman calculations of phonon excitations it is a
common\cite{pap17} practice to neglect the ${\bf A}^2$ term in the
electron-light interaction Hamiltonian (as in Eq.~\ref{eq:e-light}).
However, Ref.~\onlinecite{Basko2009} claims that under certain
conditions ${\bf A}^2$ terms are important for the Raman G peak
process.  Since these conditions are not satisfied in a typical
experimental situation ($\omega_{\rm in} \sim 2$~eV, as in
Ref.~\onlinecite{kim2012}, and assuming $\gamma=0$ would lead to a
divergent G peak enhancement, see Fig.~\ref{fig:ellf}) we refer
inclusion of ${\bf A}^2$ term to the future work, as it would go
beyond the scope of this manuscript.

\subsubsection{Influence of coherence}
\label{sec:G_coh}

\begin{figure}
\centering\includegraphics{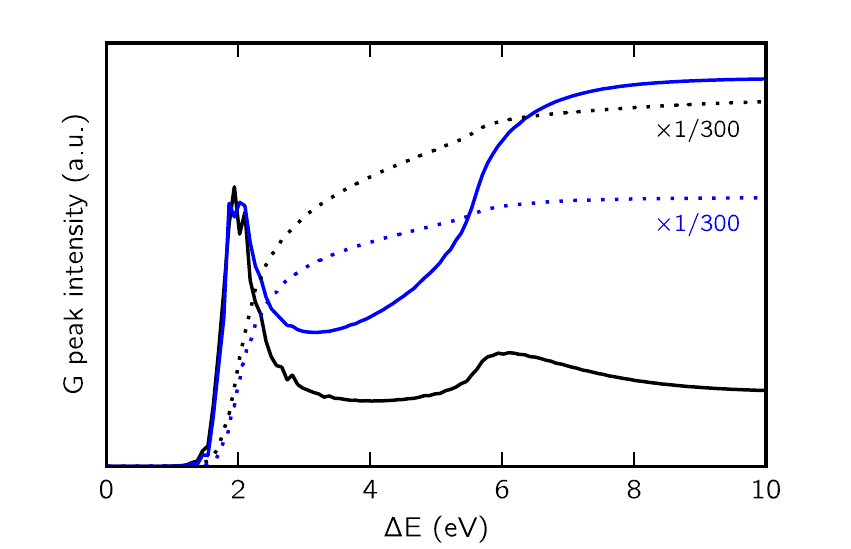}
\caption{(Color online.) Raman intensity for the G peak of a
  single-layer graphene computed in four different ways. Horizontal
  axis shows the difference in the electronic energies $\Delta E$
  appearing in the energy denominators of the Feynman diagrams as in
  Eq.~\ref{eq:example_G}. The vertical axis shows the Raman G peak
  intensity if the sum Eq.~\ref{eq:firstRaman} is performed only using
  electron-hole pairs separated in energy up to $\Delta E$.  Dotted
  lines shows the results when the sum in Eq.~\ref{eq:firstRaman} is
  performed incoherently over electron and hole states. Both dotted
  lines are downscaled 300 times in intensity. Solid lines show the
  results when the sum is performed coherently.  Blue (gray) lines
  show results when the sum is performed only over two Feynman
  diagrams shown in Fig.~\ref{fig:diag_G}~(a) and (b), while black
  lines shows results when the sum is performed over all twelve
  Feynman diagrams.  Comparing solid black line to other three lines,
  we see influence of the coherence in the electronic sum in
  Eq.~\ref{eq:firstRaman}, influence of all twelve Feynman diagrams
  from Fig.~\ref{fig:diag_G}, and influence of performing the sum up
  to energies larger than the incoming photon energy $\omega_{\rm
    in}$, (in this calculation $\omega_{\rm in}=$1.96~eV).}
\label{fig:conv}
\end{figure}

We find a large influence of coherence (interference) in the
calculation of the Raman G peak. (Similar observation was found in
Ref.~\onlinecite{Basko2009}.)  This is true both for the coherence
between different Feynman diagrams (shown in Fig.~\ref{fig:diag_G})
and for the coherence between different electronic states appearing in
Eq.~\ref{eq:firstRaman}.  Influence of both of these coherences is
illustrated in Fig.~\ref{fig:conv}. Figure \ref{fig:conv} shows four
different ways the sum given in Eq.~\ref{eq:firstRaman} is performed.
Horizontal axis of Fig.~\ref{fig:conv} shows the difference in
electronic energies ($\Delta E$) appearing in the energy denominator
as in Eq.~\ref{eq:example_G}.  The vertical axis of
Fig.~\ref{fig:conv} shows value of the Raman G peak intensity, if the
sum in Eq.~\ref{eq:firstRaman} is performed over all pairs of
electronic states with energy separation up to $\Delta E$.  Dotted
lines in the Fig.~\ref{fig:conv} show the Raman intensity for the
Raman G peak if the coherent sum appearing in Eq.~\ref{eq:firstRaman},
$\left| \sum_{{\rm A} {\rm B}} K_{A B}^{\nu} \right|^2$, is replaced
with an incoherent sum $\left( \sum_{{\rm A} {\rm B}} \left| K_{A
      B}^{\nu} \right| \right)^2$.  Solid lines show results for when
the sum is performed coherently, as in Eq.~\ref{eq:firstRaman}.
Additionally, dotted lines are downscaled 300 times in
Fig.~\ref{fig:conv} so that they can be compared more easily to the
coherent result. Blue (gray) lines in Fig.~\ref{fig:conv} represent
Raman G peak intensity when the sum in Eq.~\ref{eq:firstRaman} is
performed only over two Feynman diagrams shown in
Fig.~\ref{fig:diag_G}~(a) and (b), while black lines shows results for
all twelve first-order diagrams in Fig.~\ref{fig:diag_G}.

From Fig.~\ref{fig:conv} we can reach several conclusions about the
Raman G peak in graphene. First, we find that the coherence in
Eq.~\ref{eq:firstRaman} between different electronic states leads to
the suppression of the Raman G peak intensity by more than 300 times.
Second, in order to achieve the fully converged result, we find that
the sum in Eq.~\ref{eq:firstRaman} has to be performed over
electron-hole pairs separated in energy more than the incoming photon
energy $\omega_{\rm in}$ (1.96~eV in the case of Fig.~\ref{fig:conv}).
This is especially true for the coherent calculation (solid lines).
Third, we find that if the sum in Eq.~\ref{eq:firstRaman} is performed
only up to the energies close to the incoming photon energy
$\omega_{\rm in}$, that the sum is dominated by two diagrams shown in
Fig.~\ref{fig:diag_G}~(a) and (b).  However, if the sum in
Eq.~\ref{eq:firstRaman} is continued to the energies larger than
$\omega_{\rm in}$, Feynman diagrams from Fig.~\ref{fig:diag_G}~(c) to
(l) start to dominate (compare solid blue and black lines in
Fig.~\ref{fig:conv}).

\subsection{Raman 2D peak}
\label{sec:2D}

Similarly as in the case of the Raman G peak, we expect to see changes
in the Raman 2D peak when the angle $\ang$ is close to the critical
angle. In fact we find an even more complicated dependence of the 2D
Raman peak on angle $\ang$ than that of the Raman G peak.

Comparing the first-order Raman calculation (as for the Raman G peak)
given in Eq.~\ref{eq:firstRaman} to the second-order Raman calculation
(as for the 2D peak) given in Eq.~\ref{eq:secondRaman} we see that in
the latter case the sum is performed over all phonon momenta $\q$ in
the entire phonon Brillouin zone. Phonons at different momenta $\q$
have different frequency, $\omega_{\q}^{\nu}$, which in general would
lead to the Gaussian-like spread in the Raman intensity
$I_2(\omega_{\rm out})$, even if the phonon lifetime is infinite. This
observation is not true for the Raman G peak since it involves only
a single phonon frequency $\omega_{\bm 0}^{\nu}$, and therefore its
Lorentzian-type width comes solely from the finite phonon lifetime,
and its peak position is determined by $\omega_{\bm 0}^{\nu}$.

\begin{figure}[!t]
\centering\includegraphics{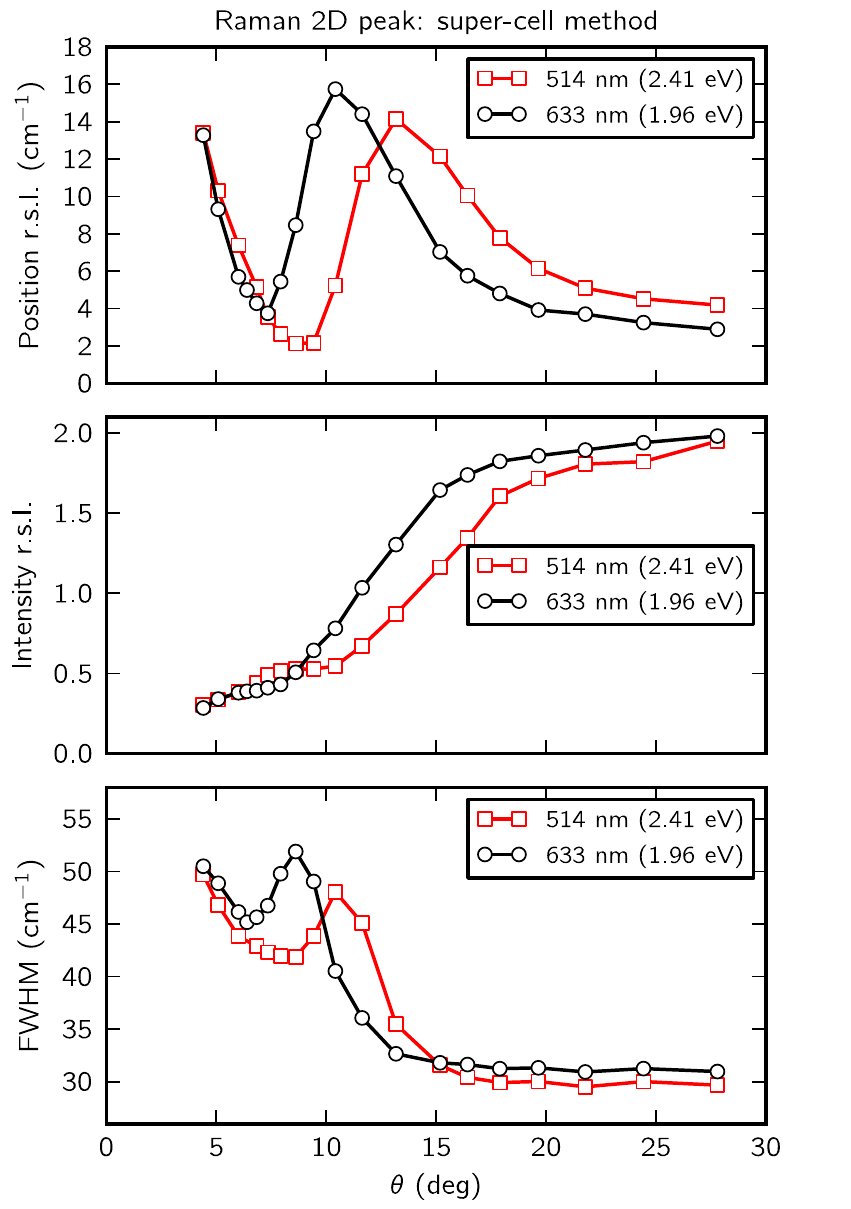}
\caption{(Color online.) Super-cell tight-binding model calculated
  position, intensity, and width of the Raman 2D peak. Black line
  indicates results for the incoming photon energy of 1.96~eV while
  the red (gray) line shows results for the incoming photon energy of
  2.41~eV.  Horizontal axis gives angle $\ang$ of the rotated
  double-layer graphene. The range of angle $\ang$ shown is from
  $0^{\circ}$ to $30^{\circ}$. For range $30^{\circ}<\ang<60^{\circ}$
  we find almost the same Raman G peak intensity for
  $\ang=30^{\circ}+\Delta$ as for the $\ang=30^{\circ}-\Delta$ case.
  To be consistent with Ref.~\onlinecite{kim2012} and other
  experimental work, fit was performed to the Lorentzian function.
  Similar results (especially for the position and the intensity) are
  obtained by a fit to the Gaussian function, see black line in
  Fig.~\ref{fig:two_gauss}.  Intensity is defined as the area under
  the peak (not peak height).  Width is defined as the full width at
  half of the peak maximum (FWHM). Peak intensity and peak position
  are defined relative to a single-layer graphene. See main text for
  more details.  Comparison with experimental data (in good agreement
  with our calculation) is shown in Ref.~\onlinecite{kim2012}.}
\label{fig:2D_tb}
\end{figure}

The super-cell tight-binding method computed Raman 2D peak position,
intensity, and width in the rotated double-layer graphene are given in
Fig.~\ref{fig:2D_tb}. For all three features of the 2D peak we find a
complex variation as a function of the angle $\ang$, especially so
near the critical angle ($\sim 10^{\circ}$ for 1.96~eV incoming
photon energy). Similarly as in the case of the Raman G peak, we find
that these features shift to the larger angle $\ang$ if the incoming
photon energy is increased.  Again, as in the case of the Raman G
peak, this behavior is consistent with the angle $\ang$ dependent
position of the Van Hove singularities shown in Fig.~\ref{fig:dos}.

The position of the Raman 2D peak in Fig.~\ref{fig:2D_tb} is indicated
relative to the single-layer graphene case at the same incoming photon
energy (since even for the single-layer case the Raman 2D peak
position depends on the incoming photon energy). We find that the
position of the Raman 2D peak of the rotated double-layer graphene is
shifted to the larger energies with respect to the single-layer
graphene case. The observed shift is non-monotonic, starting out small
($\sim 4$~cm$^{-1}$) at large angles ($>20^{\circ}$). Close to the
critical angle $\sim 10^{\circ}$ (for 1.96~eV incoming photon energy)
the shift in the peak position increases to $\sim 14$~cm$^{-1}$ and is
followed by a steep drop to $\sim 4$~cm$^{-1}$ at about 7$^{\circ}$.
At even lower angles ($<7^{\circ}$) there is a steep rise in the 2D
peak position.

The intensity of the Raman 2D peak is somewhat less complicated than the
peak position and the peak width. The Raman 2D peak intensity shows
almost a step-like change close to the critical angle $\sim
10^{\circ}$ (for 1.96 eV incoming photon energy), having an intensity
comparable to two independent single-layers at higher angles, and
$\sim 4$ times smaller intensity at the smaller angles.

The width of the Raman 2D peak at angles above 15$^{\circ}$ (for 1.96~eV
incoming photon energy) is comparable to that of a single-layer
graphene, $\sim 31$~cm$^{-1}$. At the smaller angles ($<15^{\circ}$)
there is a sharp increase in the Raman 2D peak width.  Additionally,
close to $8^{\circ}$ Raman 2D peak width suddenly jumps to
52~cm$^{-1}$. Below $8^{\circ}$ there is again a non-monotonic
behavior of the width, starting with a decrease followed by a sharp
increase below $6^{\circ}$.

\begin{figure}[!t]
\centering\includegraphics{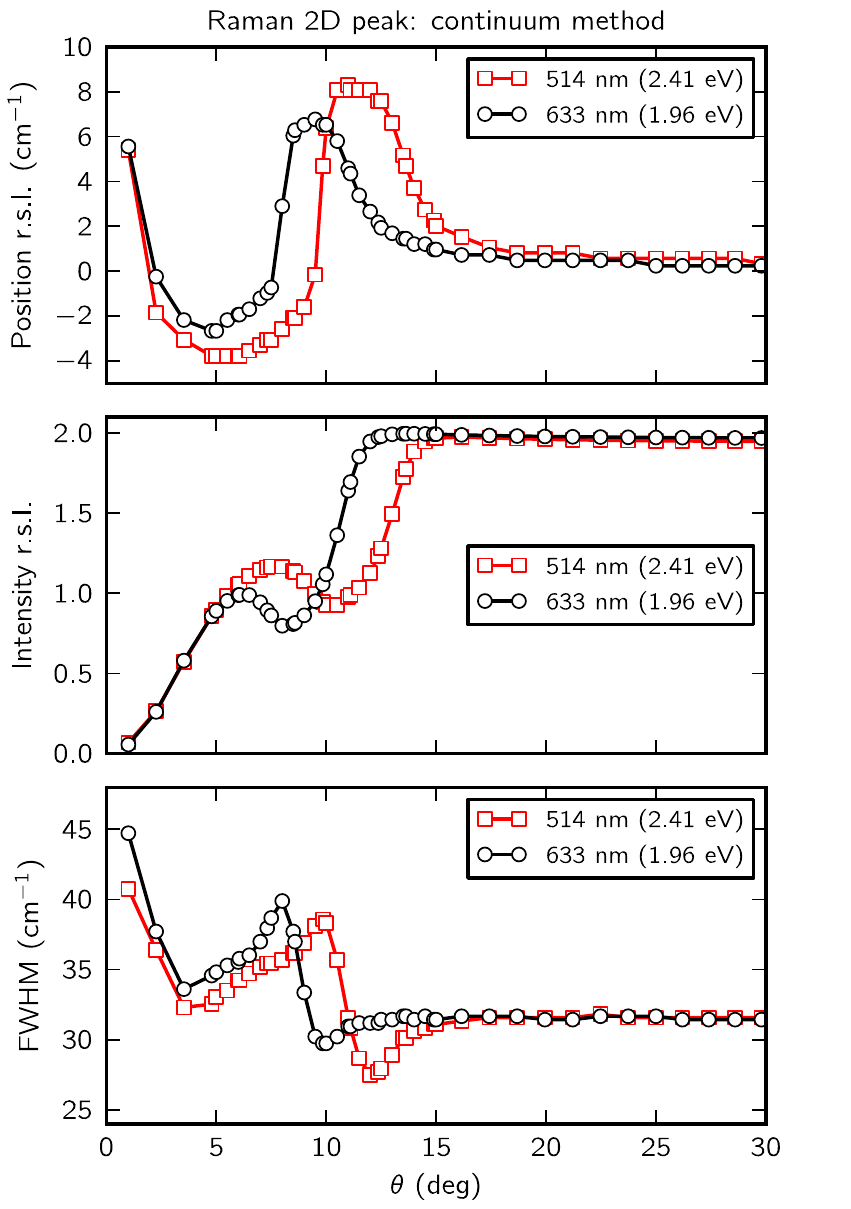}
\caption{(Color online.) Position, intensity, and width of the Raman
  2D peak using a more approximate method (continuum model).
  Conventions are the same as in Fig.~\ref{fig:2D_tb}, but the range
  of vertical scales is not the same as in Fig.~\ref{fig:2D_tb}.}
\label{fig:2D_cont}
\end{figure}

The results of the continuum model calculation of the Raman 2D peak in
Fig.~\ref{fig:2D_cont} show similar overall features as the
super-cell tight-binding calculations. The angle dependence of the
peak position, intensity and width follow the same trends in both
calculations, but the numerical values are somewhat different. In
addition, there are some spurious features present in
Fig.~\ref{fig:2D_cont} that are not present in the super-cell
tight-binding calculation. For example, the Raman 2D peak intensity
and width in the region from $\ang=5^{\circ}$ to $\ang=15^{\circ}$
show some small features not present in the super-cell tight-binding
calculation. We expect that these differences are occurring due to the
approximations introduced into the continuum model calculation (see
Sec.~\ref{sec:cont}). In particular, the lack of the trigonal warping in
the continuum model becomes especially important at large energy
of the incoming photons and for large angle $\ang$. Additionally,
the reduction of the Hilbert space becomes more important at low
angles $\ang$.

\subsubsection{Procedure to experimentally determine angle $\ang$}
\label{sec:exp_ang}

The dependence of the position, the intensity, and the width of the
Raman 2D peak on the angle $\ang$ provide a simple route to
experimentally determine angle $\ang$. However, for some range of
values of $\ang$ the position and the width of the Raman 2D peak
depend non-monotonically on the angle $\ang$. Naively, one would
expect that this would make it impossible to uniquely determine $\ang$
in that range of angles. Nevertheless, combining all three properties
of the Raman 2D peak (position, intensity, and width) make it easier
to uniquely assign angle $\ang$. Furthermore, combining Raman
measurements at two different incoming photon energies gives
additional way to uniquely determine the angle $\ang$ even in the
region where the position and the width of the Raman 2D peak depend
non-monotonically on $\ang$.  For example, if one measures for the
incoming photon energy of 1.96~eV change in the Raman 2D peak position
of 8~cm$^{-1}$, according to the black line in Fig.~\ref{fig:2D_tb}
this measurement can correspond to angle $\ang$ of $\sim5^{\circ}$,
$\sim10^{\circ}$, or $\sim15^{\circ}$.  However if one repeats the
measurement on the same sample with a larger incoming photon energy
(for example 2.41~eV as shown by the red line in Fig.~\ref{fig:2D_tb})
and the change in the Raman 2D peak position becomes smaller than
8~cm$^{-1}$, angle $\ang$ can be assigned uniquely to
$\sim10^{\circ}$. On the other hand, if the Raman 2D shift becomes
larger than 8~cm$^{-1}$ then $\ang$ is either $\sim5^{\circ}$ or
$\sim15^{\circ}$. Since these two angles are quite far apart (by
construction), other Raman 2D features like intensity or width can be
used to determine which of the two angles should be assigned. Similar
procedure can also be used for the non-monotonic dependence of the
Raman 2D peak width.

\subsubsection{Decomposition into contributing phonons}
\label{sec:2D_phon}

\begin{figure*}[p]
\begin{minipage}[c]{1.00\linewidth}
\centering\includegraphics{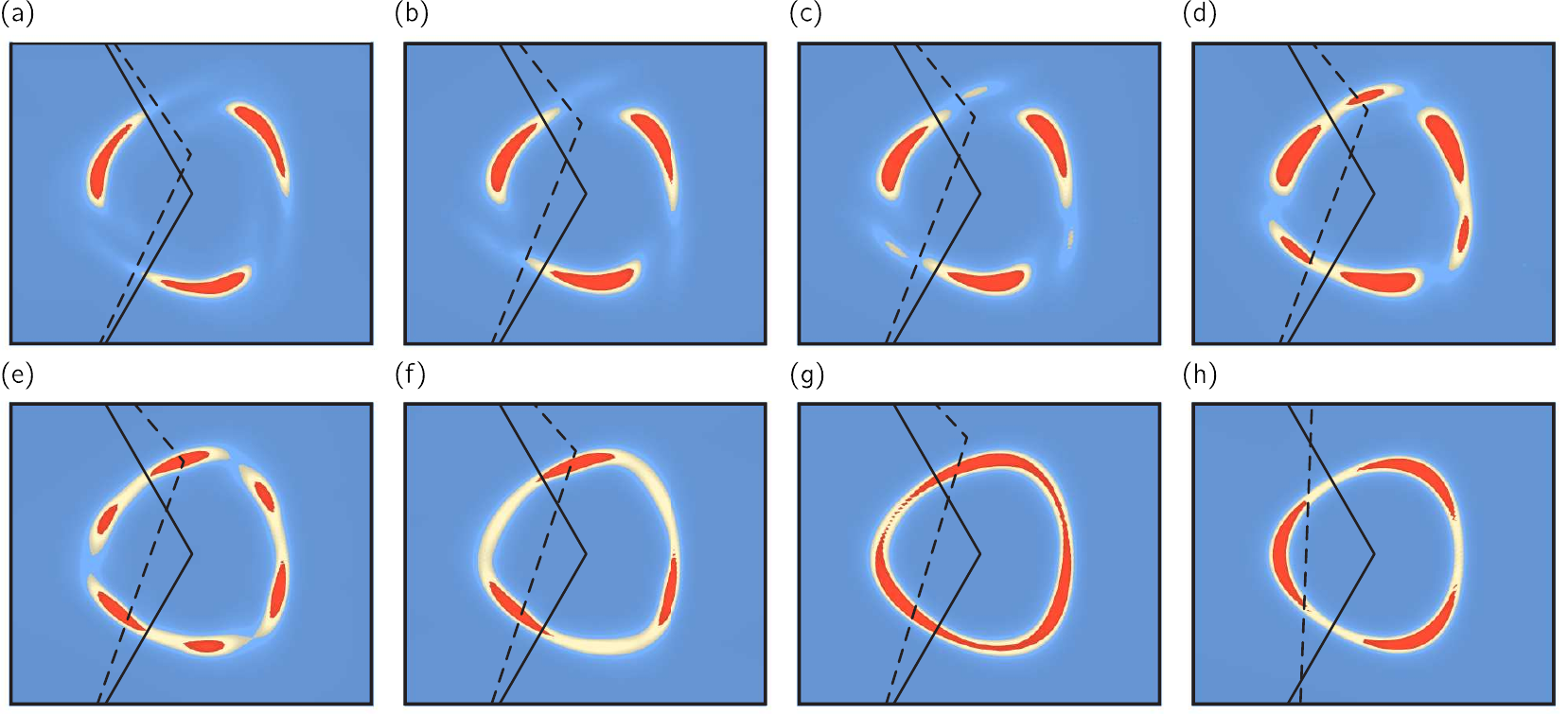}
\caption{(Color online.) Regions of the phonon Brillouin zone
  contributing the most to the Raman 2D peak, with characteristic
  triangular regions around the Brillouin zone K point. Region that
  contributes the most to the Raman 2D peak is shown in red color.
  Region of the Brillouin zone with intermediate intensity is shown in
  yellow, and that of zero intensity in blue. Color scale for each
  panel is scaled individually to the largest intensity for that
  panel, since otherwise the overall intensity of the panels for small
  angles would be too small (see Fig.~\ref{fig:2D_tb} showing decrease
  of the Raman 2D peak intensity at small angles $\ang$). The
  Brillouin zone of the bottom (solid line) and the top (dashed line)
  graphene layer are indicated.  Contributions to the 2D peak of only
  one layer (bottom) are shown for simplicity, and we only show region
  of the Brillouin zone close to the K point (approximately the same
  region is indicated with dashed line in Fig.~\ref{fig:critical}).
  The angle $\ang$ is increasing going from the panel (a) to the panel
  (h) and it equals 4.41$^{\circ}$(a), 7.93$^{\circ}$(b),
  8.61$^{\circ}$(c), 9.43$^{\circ}$(d), 10.42$^{\circ}$(e),
  11.64$^{\circ}$(f), 13.17$^{\circ}$(g), and 27.80$^{\circ}$(h).
  Calculation is performed with the incoming photon energy of 1.96~eV.
  Large transfer of weight is seen close to the critical angle in
  panel (e) when the Brillouin zone K point of the top layer is
  overlapping with the triangular region in the phonon Brillouin
  zone.}
\label{fig:phon_2D}
\end{minipage}
\\
\vspace{1.4cm}
\begin{minipage}[c]{1.00\linewidth}
 \centering\includegraphics{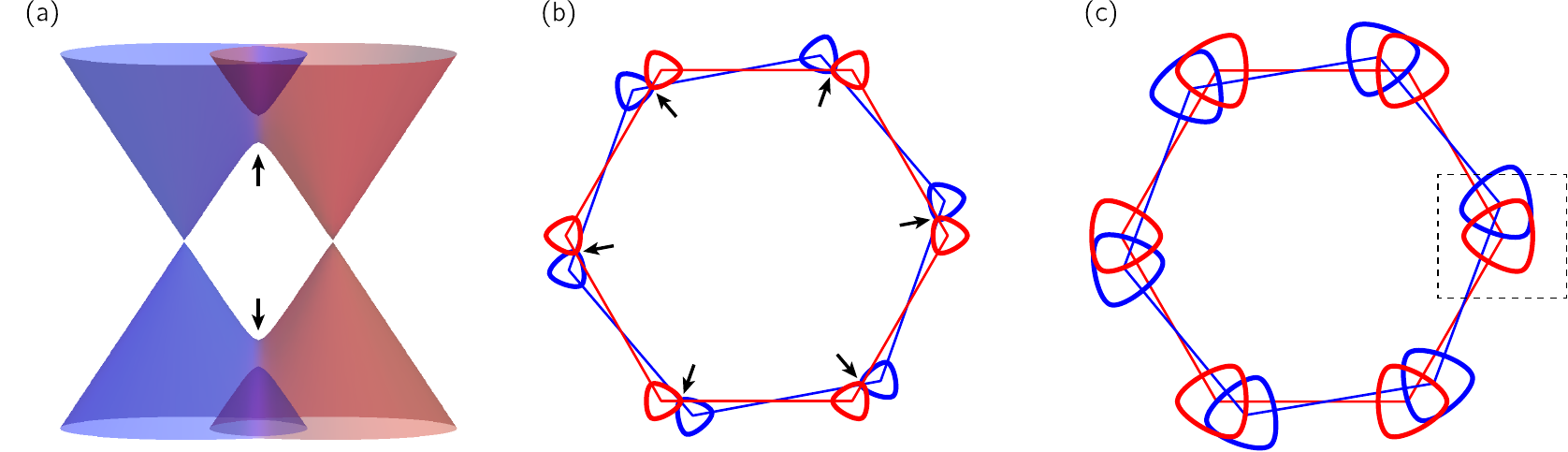}
 \caption{(Color online.) Panel (a) shows the sketch of the
   overlapping Dirac cones of the two graphene sheets (shown in red
   and blue). Dirac cones are centered at the Brillouin zone edge
   points K of the each graphene layer. Black arrows indicate the
   overlap region in which the interaction between the graphene layers
   introduces a hybridization gap in electron and hole states. Panel
   (b) shows the sketch of the isoenergy curves (for both layers) in
   the electron Brillouin zone separated in the energy by the amount
   equal to the incoming photon energy. The angle $\ang$ is close to
   the critical angle. Brillouin zone of each layer is indicated with
   red and blue hexagons. Overlap region is indicated with the black
   arrow, as in panel (a). Panel (c) shows the sketch of the nesting
   vectors in the phonon Brillouin zone connecting two Dirac cones in
   the electron Brillouin zone (corresponding to the same graphene
   layer). By construction, phonon nesting vectors have opposite
   trigonal warping to that of the electrons and are twice as far away
   from the Brillouin zone edge point K. Approximately the same region
   of the phonon Brillouin zone as in Fig.~\ref{fig:phon_2D} is
   indicated with a dashed line.}
 \label{fig:critical}
\end{minipage}
\end{figure*}

According to the Eq.~\ref{eq:secondRaman} second order Raman process
(as for the Raman 2D peak) can be decomposed into a decoherent sum of
contributions coming for the pair of phonons $(\q,\mu)$ and
$(-\q,\nu)$ with opposite momenta $\q$ and possibly different phonon
branches $\mu$ and $\nu$.  Since the phonon branches of the 2D mode
arise from the Brillouin zone folding, branch indices $\mu$ and $\nu$
can be relabeled with the rotated double-layer graphene reciprocal
vectors $\G$ (as discussed in Sec.~\ref{sec:phons}).

Figure~\ref{fig:phon_2D} shows regions of the phonon Brillouin zone
which contribute the most to the Raman 2D peak, for varying angle
$\ang$. Contributions from the phonon pair $(\q,\mu)$ and $(-\q,\nu)$
is equally distributed among the unfolded vectors $\q+\G$ with
reciprocal vector $\G$ corresponding to both $\mu$ and $\nu$.
Brillouin zone of both bottom (solid line) and top (dashed line)
single-layer graphene are indicated with black lines. For simplicity,
only contributions from phonons in one graphene layer are shown in
Fig.~\ref{fig:phon_2D}, and only the region close to the Brillouin
zone corner (K point) is shown.

For large values of angle $\ang$ [for example $\ang=27.80^{\circ}$ in
Fig.~\ref{fig:phon_2D}(h)] we find a characteristic triangular region
(red) in the phonon Brillouin zone around the K-point with the largest
contribution to the Raman 2D peak. Similar behavior we find in the
calculation of a single-layer graphene, as consistent with the
decomposition found in Ref.~\onlinecite{pap34}. At angles smaller or
equal to the critical angle, this triangular region is significantly
modified. Largest modification we find when the K-point of the
Brillouin zone of the top graphene layer is overlapping with the
triangular region in the bottom layer [see for example
Fig.~\ref{fig:phon_2D}(e)]. As shown in Fig.~\ref{fig:critical} this
modification occurs precisely at the critical angle, at which the
Dirac cones in the electron Brillouin zone are overlapping.

\subsubsection{Peak substructure, two Gaussian components of the 2D peak}
\label{sec:2D_sub}

\begin{figure}[!t]
\centering\includegraphics{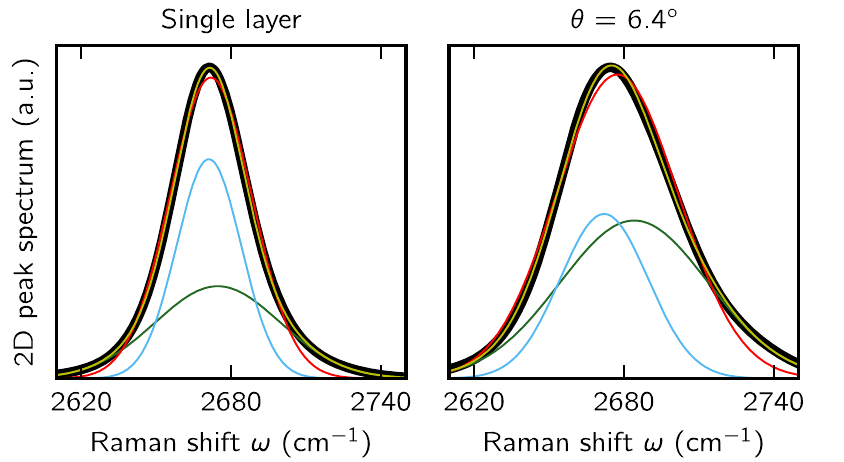}
\caption{(Color online.) Comparison of a single Gaussian fit (thin red
  line) and a two Gaussian fit (thin yellow line) of the calculated
  Raman 2D profile (thick black line) for the single layer graphene
  (left panel) and the rotated double-layer graphene with
  $\ang=6.4^{\circ}$ (right panel). Two Gaussian components of the two
  Gaussian fit are shown in green (broad component) and blue (narrow
  component). See Fig.~\ref{fig:two_gauss} for dependence of broad and
  narrow components on angle $\ang$.}
\label{fig:two_fit}
\end{figure}

Our calculations show that the profile of the Raman 2D peak
[$I_2(\omega_{\rm out})$ in Eq.~\ref{eq:secondRaman}] can be well
fitted with two Gaussians with varying position, intensity, and width
of each Gaussian function (compare black and yellow line in
Fig.~\ref{fig:two_fit}).  We find this to be true both for the
single-layer graphene and for the rotated double-layer graphene. For
the single-layer graphene importance of using two Gaussians as opposed
to only one is more subtle. However, for the rotated double-layer
graphene just below the critical angle, positions of these two
Gaussians are somewhat apart from each other, leading to the more
pronounced two-peak feature. Similar feature has been found in the
experimental measurements, near the critical angle\footnote{Private
  communication with K. Kim.}.  Furthermore, these two Gaussian
components of the Raman 2D peak behave differently as a function of
angle $\ang$ which will be of interest in analyzing angle $\ang$
dependent data for the rotated double-layer graphene.

First, let us analyze these two Gaussian components in the case of a
single-layer graphene. We find that these Gaussian components in this
case are centered around nearly the same frequency (difference is only
$3.5$~cm$^{-1}$ at 1.96~eV incoming photon energy) and have nearly the
same intensity. Additionally, we find that the width of one Gaussian
component (narrow component) is $30$~cm$^{-1}$ while the width of the
other Gaussian component (broad component) is almost two times larger,
$59$~cm$^{-1}$.

\begin{figure}[!t]
\centering\includegraphics{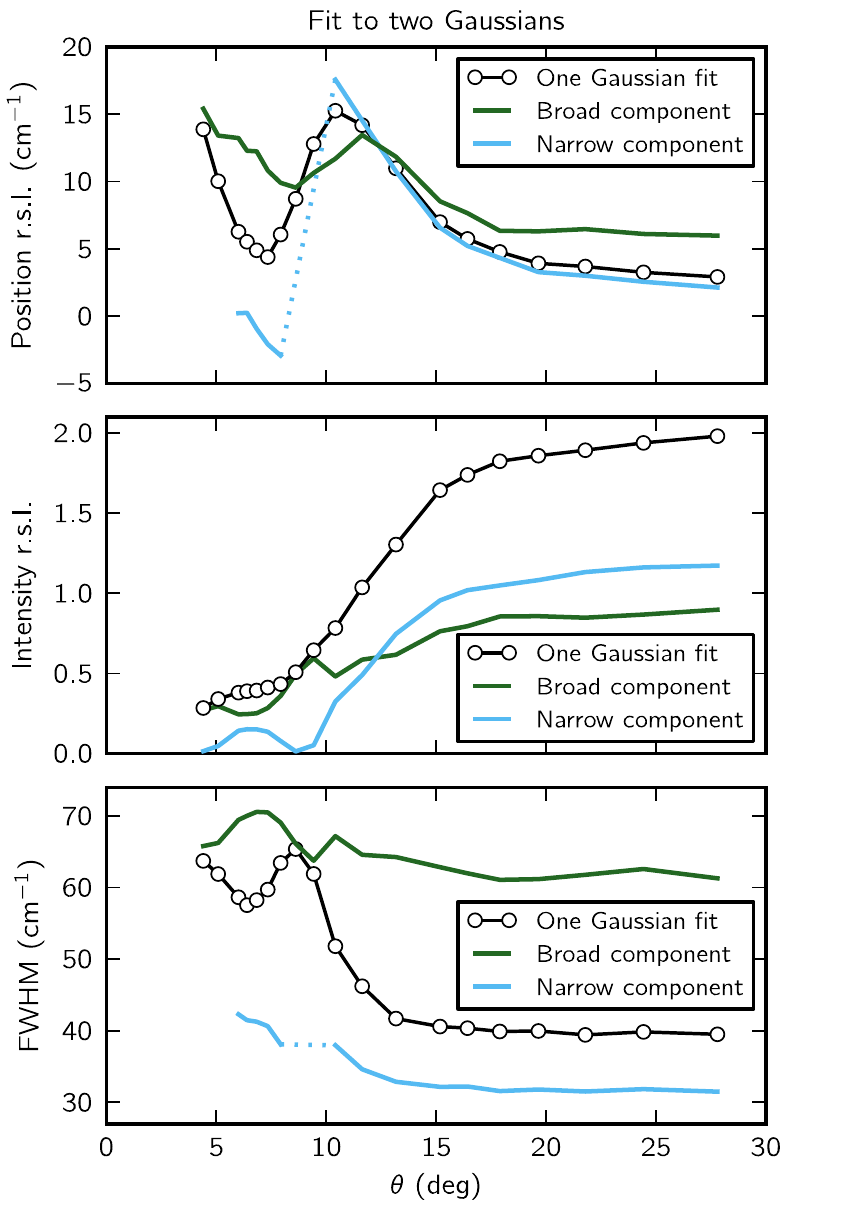}
\caption{(Color online.) Fit of the calculated Raman 2D peak to a
  single Gaussian (black line) and to two Gaussians (broad Gaussian
  component is in green and narrow in blue). Peak position and intensity
  for all three lines are given relative to the single Gaussian fit of
  the 2D Raman peak in the single-layer graphene. Other conventions
  are as in the Fig.~\ref{fig:2D_tb}. Narrow Gaussian component for some
  values of angle $\ang$ has negligible intensity, which makes fitting
  procedure ill-conditioned. For that range of angles, position and
  width of the narrow component are drawn with a straight dotted line.
  Calculation is performed for a single incoming photon energy,
  1.96~eV.}
\label{fig:two_gauss}
\end{figure}

Figure~\ref{fig:two_gauss} shows the position, the width, and the
intensity of these two Gaussian components in the case of a rotated
double-layer graphene (broad and narrow Gaussian components are shown
with different color in Fig.~\ref{fig:two_gauss}). Data in
Fig.~\ref{fig:two_gauss} is shown for the super-cell tight-binding
calculation, but similar results are obtained with the continuum
model. 

Quite surprisingly, we find that the broad Gaussian component of the
Raman 2D peak in the rotated double layer graphene is nearly
independent of the angle $\theta$. There is an overall decrease in
the intensity of the broad component below the critical angle
($\sim10^{\circ}$) but the changes in the position and the width are
almost negligible.

For the narrow Gaussian component in the rotated double layer graphene
we again find that its width almost does not depend on the angle
$\ang$. On the other hand, the peak intensity and the peak position of
the narrow component show a drastic change below the critical angle
($\sim10^{\circ}$). In particular, exactly at the critical angle the
narrow component nearly vanishes. Below the critical angle ($5^{\circ}
< \ang < 10^{\circ}$) the narrow component reappears but with
significantly lower peak position ($-3$~cm$^{-1}$ below the critical
angle as compared to 18~cm$^{-1}$ above the critical angle). At the
even lower angle ($\ang<5^{\circ}$) the narrow component nearly
disappears once again.

This appearance and disappearance of the narrow component gives an
insight into the complex behavior of the overall position, intensity,
and width of the Raman 2D peak (black line in
Fig.~\ref{fig:two_gauss}).  For example, the overall increase in the
width of the Raman 2D peak near the critical angle ($\sim10^{\circ}$)
can be explained by the disappearance of the narrow Gaussian component
at the same angle.  Similarly, reappearance of the narrow component with
lower frequency below the critical angle ($5^{\circ} < \ang <
10^{\circ}$) explains the overall change in the peak position of the
Raman 2D peak.  Additionally, reappearance of the narrow component at
the lower frequency than the broad component is consistent with the
experimentally observed two peak structure of the Raman 2D peak in the
same range of angles $\ang$.

It is tempting to interpret the broad and narrow Gaussian components
of the 2D peaks as coming from the corners of the triangular region
(inner phonons, Ref.~\onlinecite{pap34}) in Fig.~\ref{fig:phon_2D} and
from the triangular faces (outer phonons) respectively. Indeed,
similar two-peak feature of the Raman 2D peak has been found in
Ref.~\onlinecite{pap34}, but for significantly larger incoming photon
energies (3.8~eV). These two features of the Raman 2D peak were
denoted as 2D$^+$ (inner) and 2D$^-$ (outer) in Fig.~26 of
Ref.~\onlinecite{pap34}. However, origin of the two peak features we
find here is {\it decidedly} different. We demonstrate this by taking
our single-layer graphene calculation and considering only small
slices (in certain region of angles around the K-point) of the
triangular regions in the phonon Brillouin zone either near the
triangular corners or faces.  We find in both cases that the
two-Gaussian peak feature persists, with similar fitting parameters.

Instead, we find that this two-peak structure of the Raman 2D peak
originates from the sum over electron-hole pair states in
Eq.~\ref{eq:secondRaman} (not different phonon states as for the
feature found in Ref.~\onlinecite{pap34}). In particular, we find that
the electron-hole pairs which are separated by the energy close to the
incoming light energy give rise to the narrow component of the 2D peak
from Fig.~\ref{fig:two_gauss}, while the higher energy electron-hole
pairs give rise to the broad component from Fig.~\ref{fig:two_gauss}.
More specifically, for the incoming photon energy of $1.96$~eV, we
find that the narrow component of the 2D peak originates from the
electron-hole pairs separated up to $\sim2.1$~eV. Electron-hole pairs
between $\sim2.1$ and $\sim2.6$~eV give rise to the broad component.

\subsubsection{Influence of interlayer interaction on electron
  wavefunctions and eigenenergies}
\label{sec:2D_ele}

\begin{figure}[!t]
\centering\includegraphics{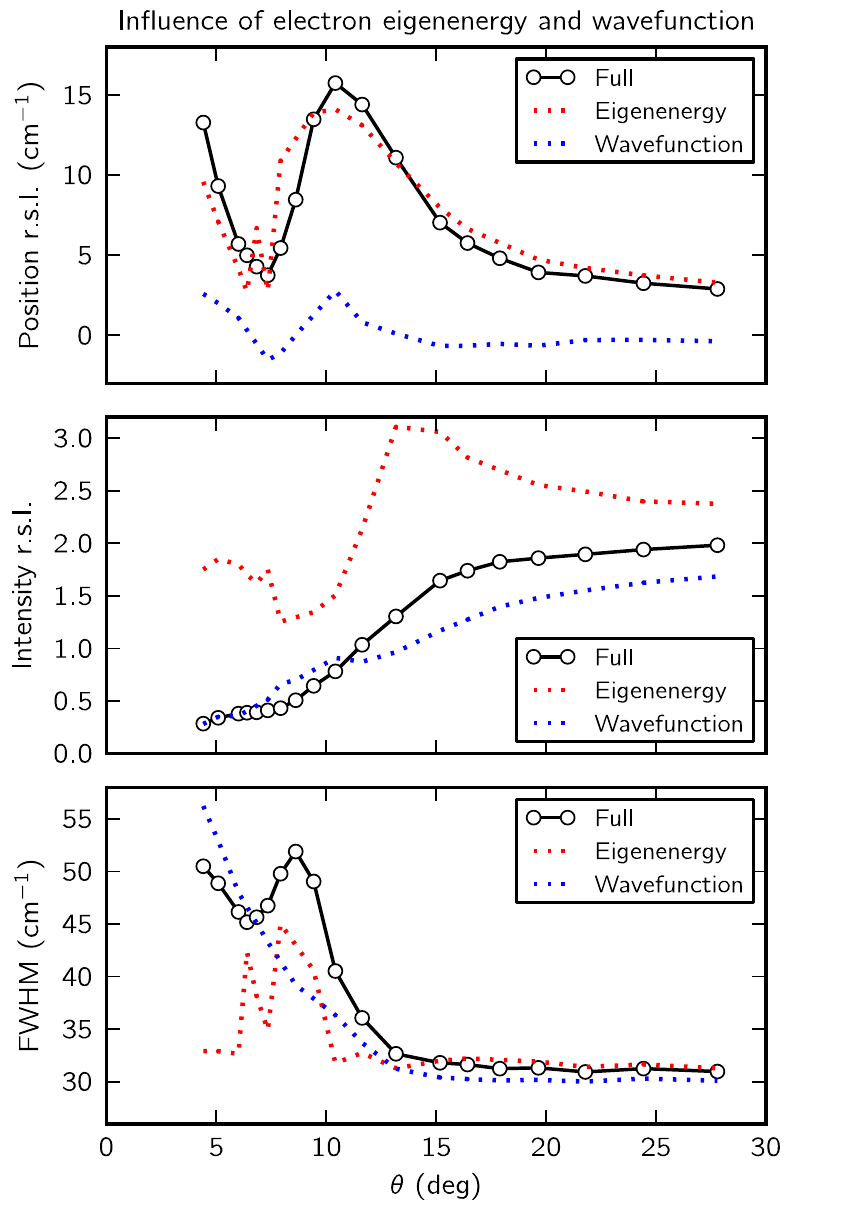}
\caption{Calculated position, intensity, and width of the Raman 2D
  peak for the incoming photon energy of 1.96~eV. Dotted lines show
  results of the calculation in which the influence of the electron
  hopping terms between two graphene layers affects only electron
  eigenenergies (red) or only electron wavefunctions (blue). See main
  text for more details.  Other conventions are the same as in
  Fig.~\ref{fig:2D_tb}.}
\label{fig:2D_ene_wf}
\end{figure}

The tight-binding model of the rotated double-layer graphene used in
our study is based on a Slater-Koster parametrization from
Ref.~\onlinecite{pap03}. This parametrization assigns a hopping term
to any pair of p$_{\rm z}$ orbitals on two carbon atoms. These carbon
atoms can either be in the same, or two different graphene layers. 
Therefore, if we set to zero all hopping terms between pair of carbon
atoms in the different graphene layers (interlayer hopping), we can
effectively turn off the interaction between the two graphene layers.

The effect of allowing the electron interlayer hopping in our
calculation is twofold.  Firstly, it affects electron wavefunctions.
The change in the electron wavefunctions modifies electron-light and
electron-phonon matrix elements, which in turn changes Raman intensity
of both G and 2D peak, as given for example in the numerators of
Eqs.~\ref{eq:example_G} and \ref{eq:example_2D}.  Secondly, interlayer
hopping affects electron eigenenergies.  Electron eigenenergies in
turn affect Raman G and 2D intensities through the denominators in for
example Eqs.~\ref{eq:example_G} and \ref{eq:example_2D}.

Figure~\ref{fig:2D_ene_wf} shows which features of the Raman 2D peak
can be explained solely by the influence of the interlayer hopping on
the electron wavefunctions, and which by the influence on the electron
eigenenergies. Dotted red (blue) line in Fig.~\ref{fig:2D_ene_wf}
shows the Raman 2D peak position, intensity, and width for the
calculation in which the interlayer hopping is given only for the
electron eigenenergies (electron wavefunctions). Solid black line in
these graphs are the same as in Fig.~\ref{fig:2D_tb}, showing the
results of the full Raman 2D peak calculation (with interlayer hopping
considered both for electron eigenenergies and wavefunctions).

From Fig.~\ref{fig:2D_ene_wf} we conclude that the position of the
Raman 2D peak is almost completely determined by the influence of the
interlayer hopping on the electron eigenergies. On the other hand,
intensity of the Raman 2D peak is determined by the interlayer hopping
influence on the electron wavefunctions. Finally, increase in the
width of the Raman 2D peak at low angles $\ang$ is well described by
the influence of the interlayer hopping on the electron wavefunctions.
However, influence of the interlayer hopping on the electron
wavefunctions does not reproduce feature in the Raman 2D peak width
near the critical angle ($\sim 10^{\circ}$).

\subsection{Limit of small and limit of large angles}
\label{sec:limit}

Here we discuss properties of the Raman 2D and G peaks of the rotated
double-layer graphene in the limit of small (close to $0^{\circ}$) and
large (close to $30^{\circ}$) angles $\ang$. For the Raman G peak we
find that in both limits ($0^{\circ}$ and $30^{\circ}$) intensity of
the G peak is similar to that of a single-layer graphene (multiplied
with number of layers in the rotated double-layer graphene, two). In
fact, for the entire range of angles $\ang$, except close to the
critical angle, we find that the Raman G peak intensity is similar to
that of a single-layer graphene (times two).

\begin{figure}[!t]
\centering\includegraphics{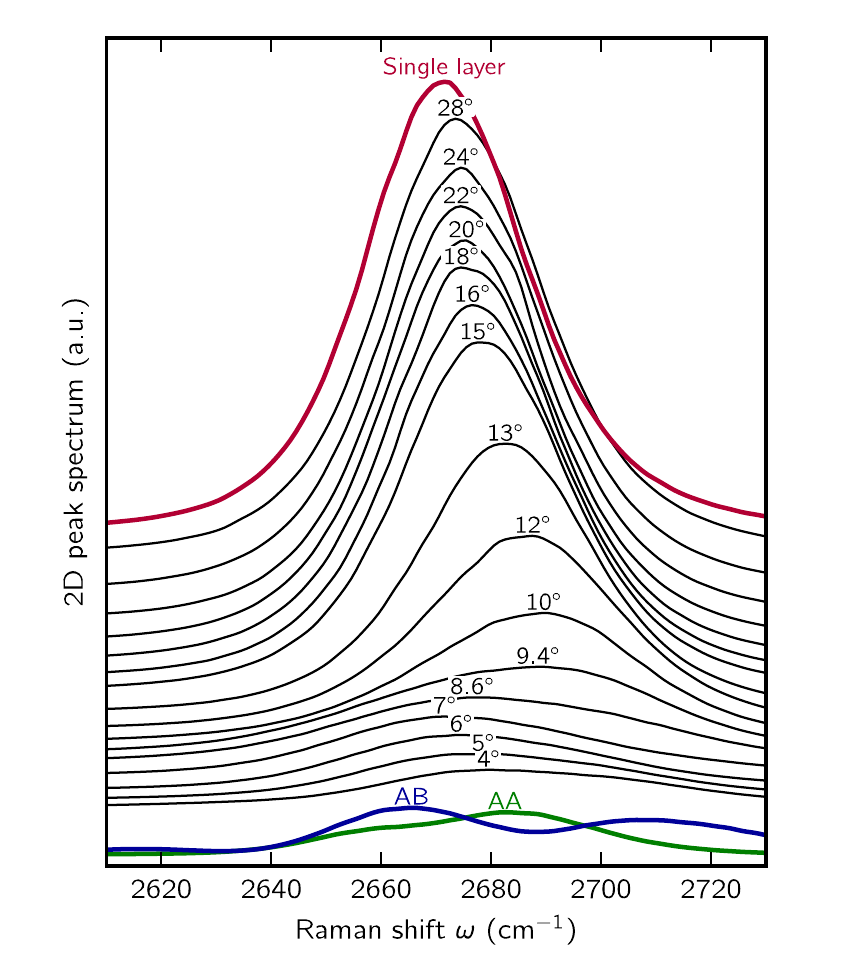}
\caption{(Color online.) Calculated Raman 2D profiles
  [$I_2(\omega_{\rm out}=\omega)$ from Eq.~\ref{eq:secondRaman}] for
  the rotated double-layer graphene (thin black lines), the
  single-layer graphene multiplied by two (thick red line), the AB
  stacked double-layer graphene (blue), and the AA stacked
  double-layer graphene (green). The rotated double-layer graphene
  spectra are shifted in the vertical direction, proportionally to the
  angle $\ang$, for clarity. Raman 2D profile of the single-layer
  graphene (red) is shifted vertically proportional to
  $\ang=30^{\circ}$.}
\label{fig:spec}
\end{figure}

The situation with the Raman 2D peak is again more complicated.
Figure~\ref{fig:spec} shows calculated Raman 2D profiles for the
rotated double-layer graphene (black) shifted for clarity in the
vertical direction proportionally to the value of the angle $\ang$.
The Raman 2D profile of the single-layer graphene (multiplied by two)
is indicated with thicker red line in Fig.~\ref{fig:spec}.  From
Fig.~\ref{fig:spec} one can see that the Raman 2D spectrum of the
rotated double-layer graphene above $\ang \approx 15^{\circ}$ is
already converging towards that of a single-layer graphene (red).

On the other hand, in the limit of a small angle $\ang$ (close to
$0^{\circ}$) Raman 2D peak intensity of the rotated double-layer
graphene is significantly smaller than that at the larger angles, or
that of the single-layer graphene. We find similar reduction in
intensity in the case of the AB (blue in Fig.~\ref{fig:spec}) and the
AA (green in Fig.~\ref{fig:spec}) stacked double-layer graphene.
Additionally, peak position and width for small angles $\ang$ are
qualitatively similar to that of the AB and AA stacked double-layer
graphene. Similarity with the AB and AA stacked double-layer graphene
is not unexpected since the rotated double-layer graphene in the limit
of very small angles $\ang$ is composed of a hexagonal super-periodic
arrangements of AB and AA stacked regions. This pattern is already
visible to some degree on Fig.~\ref{fig:unit_cell}(b) for the case of
$\ang=9.43^{\circ}$ and is even more prominent at smaller angles
$\ang$.

However, in the sharp contrast to the AB and AA stacked double-layer
graphene, we find no prominent multi-peak structure in the case of the
rotated double-layer graphene in the limit of a very small angle
$\ang$.  Furthermore, double peak structure discussed earlier in
Sec.~\ref{sec:2D_sub} is of a different origin, and separation in
frequency between the two Gaussian components is much smaller.

\section{Summary and outlook}
\label{sec:summary}

In this work we provided a theoretical description of the two most
prominent Raman signals in rotated double-layer graphene (G peak and
2D peak). We find a relatively simple dependence of the Raman G peak
intensity on the angle $\ang$. On the other hand, position, intensity,
and width of the Raman 2D peak as a function of angle $\ang$ is much
more complex. All of our findings are in good agreement with available
experimental data\cite{kim2012}. We trace the origin of the complex
dependence of the Raman 2D peak signal on the angle $\ang$ by
decomposing the Raman 2D peak into two Gaussian components with quite
different widths that are nearly independent on the angle $\ang$. In
fact, strong dependence of the intensity and position of one of the
components is responsible for the overall changes to the Raman 2D
peak.

Additionally, we discuss importance of coherence in the Raman G
peak calculation. We analyze both coherence over the various
electron-hole pairs, and coherence over the various Feynman diagrams
contributing to the Raman G peak. In the case of the Raman 2D peak we
analyze regions of the phonon Brillouin zone contributing to the Raman
signal, and explore the influence of the interlayer interaction on the
electron wavefunctions and eigenenergies.

Our study provides a way to experimentally determine angle $\ang$ of
the rotated double-layer graphene using only the Raman spectroscopy
measurement. Angle determination becomes even more robust if one
repeats Raman spectroscopy measurement with a different incoming
photon energy, as discussed in Sec.~\ref{sec:exp_ang}. Finally, this
work provides an insight into the coupling between the mechanical
degree of freedom (angle $\ang$) and the electronic degrees of freedom
(singularities in the density of states) in the rotated double-layer
graphene. We expect similar effects to occur if even more layers of
graphene are stacked on top of each other, or if different
graphene-like two-dimensional systems are stacked on top of each
other.

\begin{acknowledgments}
  We thank Gregory Samsonidze for discussion and Francesco Mauri for
  sharing data on the calculated monolayer graphene phonon band
  structure.  This work was supported by the Director, Office of
  Science, Office of Basic Energy Sciences, Materials Sciences and
  Engineering Division, U.S. Department of Energy under Contract No.
  DE-AC02-05CH11231 and by the National Science Foundation under grant
  No. DMR10-1006184 which provided for continuum model calculations.
  SGL acknowledges support of a Simons Foundation Fellowship in
  Theoretical Physics.  Computational resources were provided by the
  National Energy Research Scientific Computing Center, which is
  supported by the Office of Science of the U.S. Department of Energy.
\end{acknowledgments}

\bibliography{pap}

\end{document}